\documentclass[12pt]{article}
\usepackage{amsmath,amssymb,amsthm,amscd}
\usepackage{graphicx}
\usepackage{fullpage}
\usepackage[all]{xy}
\usepackage{axodraw}

 \makeatletter
\newfont{\footsc}{cmcsc10 at 8truept}
\newfont{\footbf}{cmbx10 at 8truept}
\newfont{\footrm}{cmr10 at 10truept}
\renewcommand{\ps@plain}{%
\renewcommand{\@oddfoot}{\footsc Matrix Representation of Renormalization in pQFT,
{\footbf August 20, 2005} \hfil\footrm\thepage}}
\makeatother 

\newcommand{\delete}[1]{}
\newcommand{\id}{{\rm id}}          
\newcommand{\End}{\mathrm{End}}     
\newcommand{\Hom}{\mathrm{Hom}}

\makeatletter
\def\section{\@startsection{section}{1}{\z@}{-3.5ex plus -1ex minus
       -.2ex}{2.3ex plus .2ex}{\large\bf}}
\def\subsection{\@startsection{subsection}{2}{\z@}{-3.25ex plus -1ex
       minus -.2ex}{1.5ex plus .2ex}{\normalsize\bf}}
\makeatother

\newcommand{\newfish}{\parbox{1.5pc}{\begin{picture}(10,10) 
\put(1,5){\qbezier(0,-2)(8,10)(16,-2)}
\put(1,5){\qbezier(0,2)(8,-10)(16,2)}
\end{picture}}}

\newcommand{\Anewfish}{\parbox{1pc}{\begin{picture}(10,10)
\put(1,5){\qbezier(-2,-8)(10,0)(-2,8)}
\put(1,5){\qbezier(2,-8)(-10,0)(2,8)}
\end{picture}}}

\newcommand{\winecup}{\parbox{1.4pc}{\begin{picture}(10,10) 
\put(5.9,-1.4){\line(2,3){9}} \put(11.1,-1.4){\line(-2,3){9}}
\put(4,7.5){\qbezier(0,1.8)(4.5,5)(9,1.8)}
\put(4,7.5){\qbezier(0,1.8)(4.5,-1.25)(9,1.8)}
\end{picture}}}

\newcommand{\roll}{\parbox{1.5pc}{\begin{picture}(10,10) 
\put(8,8){\circle{16}} \put(8,8){\qbezier(8,8)(0,0)(-8,8)}
\put(8,8){\qbezier(8,-8)(0,0)(-8,-8)}
\end{picture}}}

\newcommand{\kite}{\parbox{1.5pc}{\begin{picture}(20,15) 
\put(6.4,0){\line(-1,3){6.4}} \put(6.4,0){\line(1,3){4.8}}
\put(4.4,-1.5){\line(4,3){14}} \put(11.2,14.4){\line(2,-3){7.2}}
\put(1.6,14.4){\qbezier(0,0)(4.8,4.8)(9.6,0)}
\put(1.6,14.4){\qbezier(0,0)(4.8,-4.8)(9.6,0)}
\end{picture}}}

\def\coprod2{{\scalebox{0.13}{ 
\begin{picture}(420,150)(30,-135)
\SetWidth{3.0} \SetColor{Black} \Line(105,-60)(240,-135)
\Line(375,-60)(450,-60) \Line(240,-135)(240,15)
\Line(375,-60)(240,-135) \Line(105,-60)(240,15)
\Line(30,-60)(105,-60) \Line(240,15)(375,-60)
\end{picture}}}}

\def\l2cop{{\scalebox{0.13}{ 
\begin{picture}(195,150)(60,-135)
\SetWidth{3.0} \SetColor{Black} \Line(105,-60)(255,-135)
\Line(105,-60)(255,15) \Line(225,-115)(225,-4)
\Line(60,-60)(105,-60)
\end{picture}}}}

\def\c2cop{{\scalebox{0.13}{ 
\begin{picture}(270,90)(15,-165)
\SetWidth{3.0} \SetColor{Black} \GOval(150,-120)(45,75)(0){1.0}
\Line(75,-120)(15,-120) \Line(285,-120)(225,-120)
\end{picture}}}}

\def\r2cop{{\scalebox{0.13}{ 
\begin{picture}(195,150)(135,-135)
\SetWidth{3.0} \SetColor{Black} \Line(135,-135)(285,-60)
\Line(135,15)(285,-60) \Line(330,-60)(285,-60)
\Line(165,-115)(165,-1)
\end{picture}}}}

\def\vertex{{\scalebox{0.13}{ 
\begin{picture}(75,60)(45,-30)
\SetWidth{2.0} \SetColor{Black} \Line(45,0)(90,0)
\Line(90,0)(120,30) \Line(90,0)(120,-30)
\end{picture}}}}

\def\fishfishtree3{{\scalebox{0.2}{ 
\begin{picture}(614,166)(17,-43)
\SetWidth{1.0} \SetColor{Black}
\GBox(43.91,-26.35)(237.12,114.17){0.588}
\GBox(131.73,-17.56)(228.34,105.39){0.882}
\Line(131.73,105.39)(131.73,-17.56)
\GBox(52.69,-8.78)(149.3,96.61){0.705} \SetWidth{0.5}
\GBox(132.91,-7.61)(148.71,95.43){0.823}
\GBox(430.33,87.82)(483.03,122.95){0.588}
\GBox(544.5,87.82)(597.2,122.95){0.588} \SetWidth{1.5}
\GOval(456.68,105.39)(8.78,17.56)(0){0.588} \SetWidth{4.0}
\Line(61.48,43.91)(140.52,0) \Line(219.56,43.91)(263.47,43.91)
\Line(140.52,0)(140.52,87.82) \Line(140.52,87.82)(219.56,43.91)
\SetWidth{1.5} \Line(447.9,79.04)(447.9,17.56)
\Line(562.07,79.04)(562.07,17.56) \SetWidth{0.5}
\Vertex(447.9,17.56){5.8} \Vertex(562.07,17.56){5.8}
\SetWidth{1.5} \Line(474.24,105.39)(491.81,105.39)
\Line(421.55,105.39)(439.11,105.39)
\GOval(570.85,105.39)(8.78,17.56)(0){0.588}
\Line(588.41,105.39)(605.98,105.39)
\Line(535.72,105.39)(553.28,105.39) \SetWidth{0.5}
\GBox(390.52,-39.81)(427.99,4.68){0.705} \SetWidth{1.5}
\Line(395.2,-17.56)(377.64,-17.56)
\Line(395.2,-17.56)(439.11,8.78)
\Line(421.55,-32.79)(421.55,-2.93)
\Line(395.2,-17.56)(439.11,-43.91) \SetWidth{0.5}
\GBox(583.14,-38.64)(621.2,4.68){0.882} \SetWidth{1.5}
\Line(632.32,-17.56)(614.76,-17.56)
\Line(570.85,-43.91)(614.76,-17.56)
\Line(570.85,8.78)(614.76,-17.56)
\Line(588.41,-33.37)(588.41,-1.76)
\SetWidth{2.5} \Line(360.07,43.91)(342.51,35.13)
\Line(360.07,43.91)(342.51,52.69) \SetWidth{4.5}
\Line(281.03,43.91)(360.07,43.91) \SetWidth{2.5}
\Line(281.03,43.91)(298.6,35.13) \Line(281.03,43.91)(298.6,52.69)
\SetWidth{0.5} \Vertex(447.9,79.04){5.8}
\Vertex(562.07,79.04){5.8} \SetWidth{4.0}
\Line(61.48,43.91)(140.52,87.82) \Line(17.56,43.91)(61.48,43.91)
\Line(219.56,43.91)(140.52,0) \SetWidth{1.0}
\Line(131.73,96.61)(131.73,-8.78)
\end{picture}}}}

\def\aaaaa{{\scalebox{0.2}{ 
\begin{picture}(614,168)(69,-41)
\SetWidth{0.5} \SetColor{Black}
\GBox(544.08,-35.34)(571.98,8.37){0.764} \SetWidth{1.0}
\GBox(97.65,-40.92)(334.82,126.49){0.647} \SetWidth{0.5}
\GBox(585.93,98.58)(641.73,126.49){0.647}
\GBox(655.68,-35.34)(683.58,8.37){0.882} \SetWidth{1.0}
\GOval(613.83,112.54)(10.23,18.6)(0){0.647}
\GBox(237.16,-26.97)(320.87,112.54){0.882}
\GBox(111.61,-26.97)(195.31,112.54){0.764} \SetWidth{3.0}
\CArc(181.36,42.78)(55.8,90,270) \CArc(251.11,42.78)(55.8,-90,90)
\Line(181.36,98.58)(251.11,98.58)
\Line(251.11,-13.02)(181.36,-13.02)
\CArc(145.09,42.78)(41.85,-90,90)
\CArc(287.38,42.78)(41.85,90,270) \Line(69.75,-13.02)(153.46,0.93)
\Line(153.46,84.63)(69.75,98.58) \Line(362.72,98.58)(292.96,84.63)
\Line(292.96,0.93)(362.72,-13.02)
\Line(390.62,42.78)(404.57,56.73)
\Line(390.62,42.78)(404.57,28.83) \SetWidth{0.5}
\Vertex(613.83,84.63){7.89} \SetWidth{1.0}
\Line(594.3,111.61)(580.35,97.65)
\Line(579.42,126.49)(593.37,112.54)
\Line(635.22,112.54)(649.17,126.49)
\Line(633.36,111.61)(647.31,97.65)
\GOval(558.03,-13.02)(14.88,9.3)(0){0.764}
\Line(557.1,-27.9)(543.15,-41.85)
\Line(558.03,-26.97)(571.98,-40.92)
\Line(544.08,15.81)(558.03,1.86) \Line(558.96,1.86)(572.91,15.81)
\GOval(669.63,-13.02)(14.88,9.3)(0){0.882}
\Line(669.63,-26.97)(683.58,-40.92)
\Line(668.7,-27.9)(654.75,-41.85) \Line(655.68,15.81)(669.63,1.86)
\Line(670.56,1.86)(684.51,15.81) \SetWidth{4.0}
\Line(390.62,42.78)(516.18,42.78) \SetWidth{3.0}
\Line(516.18,42.78)(502.22,28.83)
\Line(516.18,42.78)(502.22,56.73) \SetWidth{0.5}
\Vertex(585.93,0.93){7.89} \Vertex(641.73,0.93){7.89}
\SetWidth{1.5} \Line(613.83,84.63)(585.93,0.93)
\Line(613.83,84.63)(641.73,0.93)
\end{picture}}}}

\def\ta1{{\scalebox{0.25}{ 
\begin{picture}(12,12)(38,-38)
\SetWidth{0.5} \SetColor{Black} \Vertex(45,-33){5.66}
\end{picture}}}}

\def\tb2{{\scalebox{0.25}{ 
\begin{picture}(12,42)(38,-38)
\SetWidth{0.5} \SetColor{Black} \Vertex(45,-3){5.66}
\SetWidth{1.0} \Line(45,-3)(45,-33) \SetWidth{0.5}
\Vertex(45,-33){5.66}
\end{picture}}}}

\def\tc3{{\scalebox{0.25}{ 
\begin{picture}(12,72)(38,-38)
\SetWidth{0.5} \SetColor{Black} \Vertex(45,27){5.66}
\SetWidth{1.0} \Line(45,27)(45,-3) \SetWidth{0.5}
\Vertex(45,-33){5.66} \SetWidth{1.0} \Line(45,-3)(45,-33)
\SetWidth{0.5} \Vertex(45,-3){5.66}
\end{picture}}}}

\def\td31{{\scalebox{0.25}{ 
\begin{picture}(42,42)(23,-38)
\SetWidth{0.5} \SetColor{Black} \Vertex(45,-3){5.66}
\Vertex(30,-33){5.66} \Vertex(60,-33){5.66} \SetWidth{1.0}
\Line(45,-3)(30,-33) \Line(60,-33)(45,-3)
\end{picture}}}}

\def\te4{{\scalebox{0.25}{ 
\begin{picture}(12,102)(38,-8)
\SetWidth{0.5} \SetColor{Black} \Vertex(45,57){5.66}
\Vertex(45,-3){5.66} \Vertex(45,27){5.66} \Vertex(45,87){5.66}
\SetWidth{1.0} \Line(45,57)(45,27) \Line(45,-3)(45,27)
\Line(45,57)(45,87)
\end{picture}}}}

\def\tf41{{\scalebox{0.25}{ 
\begin{picture}(42,72)(38,-8)
\SetWidth{0.5} \SetColor{Black} \Vertex(45,27){5.66}
\Vertex(45,-3){5.66} \SetWidth{1.0} \Line(45,27)(45,-3)
\SetWidth{0.5} \Vertex(60,57){5.66} \SetWidth{1.0}
\Line(45,27)(60,57) \SetWidth{0.5} \Vertex(75,27){5.66}
\SetWidth{1.0} \Line(75,27)(60,57)
\end{picture}}}}

\def\tg42{{\scalebox{0.25}{ 
\begin{picture}(42,72)(8,-8)
\SetWidth{0.5} \SetColor{Black} \Vertex(45,27){5.66}
\Vertex(45,-3){5.66} \SetWidth{1.0} \Line(45,27)(45,-3)
\SetWidth{0.5} \Vertex(15,27){5.66} \Vertex(30,57){5.66}
\SetWidth{1.0} \Line(15,27)(30,57) \Line(45,27)(30,57)
\end{picture}}}}

\def\th43{{\scalebox{0.25}{ 
\begin{picture}(42,42)(8,-8)
\SetWidth{0.5} \SetColor{Black} \Vertex(45,-3){5.66}
\Vertex(15,-3){5.66} \Vertex(30,27){5.66} \SetWidth{1.0}
\Line(15,-3)(30,27) \Line(45,-3)(30,27) \Line(30,27)(30,-3)
\SetWidth{0.5} \Vertex(30,-3){5.66}
\end{picture}}}}

\def\thj44{{\scalebox{0.25}{ 
\begin{picture}(42,72)(8,-8)
\SetWidth{0.5} \SetColor{Black} \Vertex(30,57){5.66}
\SetWidth{1.0} \Line(30,57)(30,27) \SetWidth{0.5}
\Vertex(30,27){5.66} \SetWidth{1.0} \Line(45,-3)(30,27)
\SetWidth{0.5} \Vertex(45,-3){5.66} \Vertex(15,-3){5.66}
\SetWidth{1.0} \Line(15,-3)(30,27)
\end{picture}}}}

\def\ti5{{\scalebox{0.25}{ 
\begin{picture}(12,132)(23,-8)
\SetWidth{0.5} \SetColor{Black} \Vertex(30,117){5.66}
\SetWidth{1.0} \Line(30,117)(30,87) \SetWidth{0.5}
\Vertex(30,87){5.66} \Vertex(30,57){5.66} \Vertex(30,27){5.66}
\Vertex(30,-3){5.66} \SetWidth{1.0} \Line(30,-3)(30,27)
\Line(30,27)(30,57) \Line(30,87)(30,57)
\end{picture}}}}

\def\tj51{{\scalebox{0.25}{ 
\begin{picture}(42,102)(53,-38)
\SetWidth{0.5} \SetColor{Black} \Vertex(61,27){4.24}
\SetWidth{1.0} \Line(75,57)(90,27) \Line(60,27)(75,57)
\SetWidth{0.5} \Vertex(90,-3){5.66} \Vertex(60,27){5.66}
\Vertex(75,57){5.66} \Vertex(90,-33){5.66} \SetWidth{1.0}
\Line(90,-33)(90,-3) \Line(90,-3)(90,27) \SetWidth{0.5}
\Vertex(90,27){5.66}
\end{picture}}}}

\def\tk52{{\scalebox{0.25}{ 
\begin{picture}(42,102)(23,-8)
\SetWidth{0.5} \SetColor{Black} \Vertex(60,57){5.66}
\Vertex(45,87){5.66} \SetWidth{1.0} \Line(45,87)(60,57)
\SetWidth{0.5} \Vertex(30,57){5.66} \SetWidth{1.0}
\Line(30,57)(45,87) \SetWidth{0.5} \Vertex(30,-3){5.66}
\SetWidth{1.0} \Line(30,-3)(30,27) \SetWidth{0.5}
\Vertex(30,27){5.66} \SetWidth{1.0} \Line(30,57)(30,27)
\end{picture}}}}

\def\tl53{{\scalebox{0.25}{ 
\begin{picture}(42,102)(8,-8)
\SetWidth{0.5} \SetColor{Black} \Vertex(30,57){5.66}
\Vertex(30,27){5.66} \SetWidth{1.0} \Line(30,57)(30,27)
\SetWidth{0.5} \Vertex(30,87){5.66} \SetWidth{1.0}
\Line(30,27)(45,-3) \SetWidth{0.5} \Vertex(15,-3){5.66}
\SetWidth{1.0} \Line(15,-3)(30,27) \Line(30,57)(30,87)
\SetWidth{0.5} \Vertex(45,-3){5.66}
\end{picture}}}}

\def\tm54{{\scalebox{0.25}{ 
\begin{picture}(42,72)(8,-38)
\SetWidth{0.5} \SetColor{Black} \Vertex(30,-3){5.66}
\SetWidth{1.0} \Line(30,27)(30,-3) \Line(30,-3)(45,-33)
\SetWidth{0.5} \Vertex(15,-33){5.66} \SetWidth{1.0}
\Line(15,-33)(30,-3) \SetWidth{0.5} \Vertex(45,-33){5.66}
\SetWidth{1.0} \Line(30,-33)(30,-3) \SetWidth{0.5}
\Vertex(30,-33){5.66} \Vertex(30,27){5.66}
\end{picture}}}}

\def\tn55{{\scalebox{0.25}{ 
\begin{picture}(42,72)(8,-38)
\SetWidth{0.5} \SetColor{Black} \Vertex(15,-33){5.66}
\Vertex(45,-33){5.66} \Vertex(30,27){5.66} \SetWidth{1.0}
\Line(45,-33)(45,-3) \SetWidth{0.5} \Vertex(45,-3){5.66}
\Vertex(15,-3){5.66} \SetWidth{1.0} \Line(30,27)(45,-3)
\Line(15,-3)(30,27) \Line(15,-3)(15,-33)
\end{picture}}}}

\def\tp56{{\scalebox{0.25}{ 
\begin{picture}(66,111)(0,0)
\SetWidth{0.5} \SetColor{Black} \Vertex(30,66){5.66}
\Vertex(45,36){5.66} \SetWidth{1.0} \Line(30,66)(45,36)
\Line(15,36)(30,66) \SetWidth{0.5} \Vertex(30,6){5.66}
\Vertex(60,6){5.66} \SetWidth{1.0} \Line(60,6)(45,36)
\SetWidth{0.5}
\SetWidth{1.0} \Line(45,36)(30,6) \SetWidth{0.5}
\Vertex(15,36){5.66}
\end{picture}}}}

\def\tq57{{\scalebox{0.25}{ 
\begin{picture}(81,111)(0,0)
\SetWidth{0.5} \SetColor{Black} \Vertex(45,36){5.66}
\Vertex(30,6){5.66} \Vertex(60,6){5.66} \SetWidth{1.0}
\Line(60,6)(45,36) \SetWidth{0.5}
\SetWidth{1.0} \Line(45,36)(30,6) \SetWidth{0.5}
\Vertex(75,36){5.66} \SetWidth{1.0} \Line(45,36)(60,66)
\Line(60,66)(75,36) \SetWidth{0.5} \Vertex(60,66){5.66}
\end{picture}}}}

\def\tr58{{\scalebox{0.25}{ 
\begin{picture}(81,111)(0,0)
\SetWidth{0.5} \SetColor{Black} \Vertex(60,6){5.66}
\Vertex(75,36){5.66} \SetWidth{1.0} \Line(60,66)(75,36)
\SetWidth{0.5} \Vertex(60,66){5.66}
\SetWidth{1.0} \Line(60,36)(60,66) \Line(60,6)(60,36)
\SetWidth{0.5} \Vertex(60,36){5.66} \Vertex(45,36){5.66}
\SetWidth{1.0} \Line(60,66)(45,36)
\end{picture}}}}

\def\ts59{{\scalebox{0.25}{ 
\begin{picture}(81,111)(0,0)
\SetWidth{0.5} \SetColor{Black}
\Vertex(75,36){5.66} \SetWidth{1.0} \Line(60,66)(75,36)
\SetWidth{0.5} \Vertex(60,66){5.66}
\SetWidth{1.0} \Line(60,36)(60,66) \SetWidth{0.5}
\Vertex(60,36){5.66} \Vertex(45,36){5.66} \SetWidth{1.0}
\Line(60,66)(45,36) \Line(75,6)(75,36) \SetWidth{0.5}
\Vertex(75,6){5.66}
\end{picture}}}}

\def\tt591{{\scalebox{0.25}{ 
\begin{picture}(81,111)(0,0)
\SetWidth{0.5} \SetColor{Black}
\Vertex(75,36){5.66} \SetWidth{1.0} \Line(60,66)(75,36)
\SetWidth{0.5} \Vertex(60,66){5.66}
\SetWidth{1.0} \Line(60,36)(60,66) \SetWidth{0.5}
\Vertex(60,36){5.66} \Vertex(45,36){5.66} \SetWidth{1.0}
\Line(60,66)(45,36) \SetWidth{0.5} \Vertex(45,6){5.66}
\SetWidth{1.0} \Line(45,6)(45,36)
\end{picture}}}}

\theoremstyle{plain}
\newtheorem{thm}{Theorem}                
\newtheorem{prop}[thm]{Proposition}      
\newtheorem{lema}[thm]{Lemma}            
\newtheorem{corl}[thm]{Corollary}        
\newtheorem{rmk}[thm]{Remark}            
\newtheorem{exams}[thm]{Examples}

\theoremstyle{definition}
\newtheorem{defn}{Definition}            

\begin{document}

\title{\bf{Matrix Representation of Renormalization in
           Perturbative Quantum Field Theory}}

\author{
KURUSCH EBRAHIMI-FARD
\\
Physics Institute, Bonn University, Nussallee 12,
\\
Bonn 53115, Germany\
\\
{\small{e-mail: {\texttt{fard@th.physik.uni-bonn.de}}}}\
\\[0.2cm]and\\[0.2cm]
LI GUO
\\
Department of Mathematics and Computer Science,
Rutgers University, \\
Newark, NJ 07102, USA\
\\
{\small{e-mail: {\texttt{liguo@newark.rutgers.edu}}}}\ }

\date{ August 20, 2005}

\maketitle

\vskip 2cm


\begin{abstract}
We formulate the Hopf algebraic approach of Connes and Kreimer to
renormalization in perturbative quantum field theory using
triangular matrix representation. We give a Rota--Baxter
anti-homomorphism from general regularized functionals on the
Feynman graph Hopf algebra to triangular matrices with entries in
a Rota--Baxter algebra. For characters mapping to the group of
unipotent triangular matrices we derive the algebraic Birkhoff
decomposition for matrices using Spitzer's identity. This simple
matrix factorization is applied to characterize and calculate
perturbative renormalization.\\
\end{abstract}

\noindent {\footnotesize{${}\phantom{a}$ 2001 PACS Classification:
03.70.+k, 11.10.Gh, 02.10.Hh, 02.10.Ox}}

\noindent {\footnotesize{
\begin{tabular}{ll}
 Keywords:& \hspace{-0.3cm}Rota--Baxter operators, Atkinson's theorem,
           Spitzer's identity, Baker--Campbell--Hausdorff formula,\\
          & \hspace{-0.3cm}matrix calculus, renormalization, Hopf algebra of renormalization,
           Birkhoff decomposition\\
\end{tabular}}}

\newpage

\tableofcontents

\cleardoublepage


\section{Introduction}

Most, if not all, of the interesting and relevant 4-dimensional
quantum field theories suffer from ultraviolet divergencies and
need to be renormalized~\cite{Collins}. The basic idea of the
theory of perturbative renormalization in quantum field theory
goes back to Kramer~\cite{Brown}, and was successfully applied for
the first time in a 1947 paper by Bethe~\cite{Bethe}, dealing with
a concrete problem in perturbative quantum electrodynamics (QED).
Five decades later, Dirk Kreimer~\cite{Kreimer1} uncovered a Hopf
algebra structure underlying the intricate combinatorial-algebraic
structure of renormalization in general perturbative quantum field
theory (QFT), hereby providing a sound and useful mathematical
foundation for this most important achievement of theoretical
physics. Later, Kreimer~\cite{Kreimer2,Kreimer3,Kreimer4,Kreimer5}
and collaborators
\cite{BergbauerKreimer,BKK,KBroadhurst1,KBroadhurst2,KBroadhurst3,KBroadhurst4,KDelbourgo},
especially Connes and Kreimer~\cite{CKI,CKII,CKIII,CK4} further
developed the Hopf--algebraic approach, connecting it to
non-commutative geometry.

In their approach, one particle irreducible (1PI) Feynman graphs,
as the building blocks of perturbative QFT and renormalization,
are organized into a connected, graded, commutative,
non-cocommutative Hopf $\mathbb{C}$-algebra
$\mathcal{H}_{\mathcal{F}}$. The restricted dual of this Hopf
algebra of Feynman graphs, denoted by
$\mathcal{H}^*_{\mathcal{F}}$, contains the group
$G:=char(\mathcal{H}_{\mathcal{F}},\mathbb{C})$ of characters,
that is, algebra homomorphisms from $\mathcal{H}_{\mathcal{F}}$ to
the underlying base field $\mathbb{C}$. Feynman rules naturally
provide a special class of such characters. The group $G$ is
generated by the Lie algebra $g:=\partial
char(\mathcal{H}_{\mathcal{F}},\mathbb{C})$, formed by
derivations, or so-called infinitesimal characters. We refer the
reader to \cite{FG,FGV,Kreimer3,Manchon,Varilly} for more details.

Dealing with the ultraviolet divergencies demands a regularization
plus a renormalization scheme. As a main example for the former
serves dimensional regularization, where we replace the above base
field $\mathbb{C}$ by Laurent series
$A=\mathbb{C}[\varepsilon^{-1},\varepsilon]]$. They form equipped
with the ordinary multiplication a commutative Rota--Baxter
algebra, and we consider $G_A:=char(\mathcal{H}_{\mathcal{F}},A)$,
respectively $g_A:=\partial char(\mathcal{H}_{\mathcal{F}},A)$.
The Hopf algebra of Feynman graphs and, via the famous
Milnor--Moore Theorem \cite{FGV,MilnorMoore}, its equivalent Lie
algebra and Lie group of characters, enabled Connes--Kreimer to
capture the process of renormalization in terms of a so-called
algebraic Birkhoff decomposition of regularized Feynman rules
characters, giving rise to a link to the Riemann--Hilbert problem
\cite{CKII,CKIII}. See also~\cite{CM}.

In \cite{EGK1,EGK2} the intimate link between the notion of
Rota--Baxter algebras and the work of Connes--Kreimer was
explored. It was shown that a non-commutative generalization of a
classical result with origin in fluctuation theory of probability,
known as Spitzer's identity for Rota--Baxter algebras
\cite{Baxter,Sp}, lies at the heart of the Connes--Kreimer
decomposition theorem for regularized Hopf algebra characters.
This provides a natural way to derive Bogoliubov's recursions for
the counter term and renormalized Feynman rules. This approach
emphasizes the Lie algebra structure on Feynman graphs and the
corresponding Lie group of characters. The former is closely
related to the more general insertion and elimination Lie algebra
of Feynman graphs, which was studied in detail first by
Connes--Kreimer in \cite{CK4}, and recently in the context of
matrices by Mencattini and Kreimer
\cite{KMencattini1,KMencattini2}.

The Hopf algebras of Feynman graphs found in renormalization
theory fall into the class of so-called combinatorial Hopf
algebras. In this paper, we obtain a matrix representation of the
full space $\Hom(\mathcal{H}_\mathcal{F},A)$ of regularized
functionals, thus in particular of $G_A$ and $g_A$, directly from
the coproduct of the Hopf algebras $\mathcal{H}_\mathcal{F}$. In
fact, the matrix representation applies to a large class of Hopf
algebras $\mathcal{H}$ and gives an injective anti-homomorphism of
Rota--Baxter algebras from $\Hom(\mathcal{H},A)$ to the infinite
size upper triangular matrices $\mathcal{M}^u_\infty(A)$.
Therefore, in the particular case of QFT renormalization, i.e.,
when $\mathcal{H}=\mathcal{H}_\mathcal{F}$, the whole process of
perturbative renormalization of regularized Feynman characters,
presented in the context of Rota--Baxter algebras
\cite{EGK1,EGK2}, is translated to a parallel, but more
transparent process involving matrices. The upper (or lower)
triangular matrices with entries from the target space of such
Hopf algebra functionals form a complete filtered non-commutative
associative Rota--Baxter algebra, giving rise to Spitzer's
identity and a matrix factorization, from which we easily read off
the Birkhoff decomposition of Connes--Kreimer.

Our work was partly motivated by Berg and
Cartier's~\cite{BergCartier}, which used the pre-Lie insertion
product on Feynman graphs to investigate the renormalization Hopf
algebra of Connes and Kreimer in terms of a lower triangular
matrix representation of the Lie group $G_A$ and its generating
Lie algebra~$g_A$. We will compare with their work in
subsection~\ref{sect:Berg-Cartier}.
\smallskip

The following is a summary of the paper. In the next section we
review the notion of complete filtered Rota--Baxter algebras and
Spitzer's identity. Theorem \ref{complRBo} is a generalization of
Spitzer's classical result to associative non-commutative
Rota--Baxter algebras, giving rise to a factorization theorem with
recursively defined solutions. As an interesting and useful
application of this theorem we formulate explicitly the
factorization of unipotent triangular matrices with entries in a
commutative Rota--Baxter algebra, and provide detailed
calculations of small size matrices. Section \ref{CK-BC} deals
with the application of the matrix Rota--Baxter algebras to the
Hopf--algebraic approach to the process of renormalization in
perturbative QFT. Subsection \ref{matrixRep} is the heart to this
paper. We define a representation of the dual space of the
renormalization Hopf algebra of Connes--Kreimer in terms of upper
triangular matrices, using the coproduct structure map. This gives
rise to a (Lie) group of upper triangular matrices with unit
diagonal, representing the group of regularized Hopf algebra
characters. The associated generating Lie algebra of infinitesimal
characters, or derivations, is represented by nilpotent upper
triangular matrices. We include several explicit examples,
calculating the renormalization of amplitudes of Feynman diagrams
up to three loops in dimensionally regularized four dimensional
$\varphi^4$-theory as a quantum field theory toy model. Further
applications and calculations of our results to renormalization in
perturbative QFT are detailed in the companion article with
J.~M.~Gracia-Bond\'ia and J.~C.~V\'arilly~\cite{Echo}. We finish
with a comment on the work of Berg and Cartier giving matrix Lie
algebra representation using the pre-Lie algebra structure on
Feynman graphs.
\medskip


\section{Complete Rota--Baxter algebras of triangular matrices}

In this paper $\mathbb{K}$ denotes a field of characteristic zero
with unit denoted by $1_\mathbb{K}$ and often by $1$. All algebras
are assumed to be unital associative $\mathbb{K}$-algebras, if not
stated otherwise, with unit identified with $1_\mathbb{K}$.


\subsection{Rota--Baxter algebras}
We recall basic concepts and properties of Rota--Baxter algebras.
The reader may consult the following literature
\cite{Atkinson,Baxter,GuoKeigher,Guo1,Ro1,Ro2,Ro3} for more
information.

By a {\bf{Rota--Baxter algebra}} in this paper, we always mean a
Rota--Baxter $\mathbb{K}$-algebra of weight one, that is, a
$\mathbb{K}$-algebra $A$ with a {\bf{Rota--Baxter map}} $R:A\to
A$, fulfilling the relation
\begin{equation}
    \label{RB}
    R(x)R(y)+R(xy)=R\big(R(x)y+xR(y)\big), \; \forall x,y\in A.
\end{equation}
So we also denote a Rota--Baxter algebra by a pair $(A,R)$. For
later reference we mention the example of the well-known
dimensional regularization scheme where
$A=\mathbb{C}[\varepsilon^{-1},\varepsilon]]$, the field of
Laurent series. One easily shows that $A$ is a commutative
Rota--Baxter algebra, with the Rota--Baxter map $R:=R_{ms}$ being
the pole part projection, known under the name of minimal
subtraction (MS) scheme
$$
 R_{ms}\left( \sum_{k=-N}^{\infty} a_k\varepsilon^k \right):=
                                    \sum_{k=-N}^{-1} a_k \varepsilon^k.
$$

For a Rota--Baxter map $R$ on $A$, the map $\tilde{R}:=\id_A-R$ is
also a Rota--Baxter map. Further we have the mixed relation
\begin{equation}
    \label{mixedRB}
   R(x)\tilde{R}(y)=R\big(x\tilde{R}(y)\big) + \tilde{R}\big(R(x)y\big),\;\;
    \forall x,y \in A.
\end{equation}
The images of both Rota--Baxter maps $R$ and $\tilde{R}$ form
non-unital subalgebras of $A$. As a trivial observation we state
the fact that every algebra $A$ is a Rota--Baxter algebra with
Rota--Baxter pair $\id_A$ and $\widetilde{\id_A}=0$. A
homomorphism $f:(A,R)\to (A',R')$ between two Rota--Baxter
algebras is a ring homomorphism such that $f\circ R=R'\circ f$.
\smallskip

A {\bf{Rota--Baxter ideal}} of a Rota--Baxter algebra $(A,R)$ is
an ideal $I$ of $A$ such that $R(I)\subseteq I$.
Let $(A,R)$ be an associative Rota--Baxter algebra. The Rota--Baxter relation
extends to the Lie algebra $\mathcal{L}_{A}$ with commutator $[x,y]:=xy-yx$,
$\forall x,y \in A$. In other words,
$$
      [R(x),R(y)]+ R\big([x,y]\big) = R\big([R(x),y] + [x,R(y)]\big),
      \forall\ x,y\,\in \mathcal{L}_A,
$$
making $(\mathcal{L}_{A},R)$ into a {\bf{Rota--Baxter Lie
algebra}}.

Given a Rota--Baxter algebra $(A,R)$, we define on the
$\mathbb{K}$-vector space underlying $A$ the following so-called
{\bf{double Rota--Baxter product}}
\begin{equation}
   a *_R b :=R(a)b + a R(b) - ab,\ \forall\, a,b\, \in A. \label{double}
\end{equation}
Then the vector space $A$ equipped with the product $*_R$ and
operator $R$ is again a Rota--Baxter algebra (of weight one),
denoted by $A_R$ and called the {\bf{double Rota--Baxter algebra
of $A$}}. Further, The Rota--Baxter map $R$ becomes an algebra
homomorphism from $A_R$ to $A$:
\begin{equation}
 \label{RBhom}
        R(a *_R b)=R(a) R(b), \;\;\forall a,b \in A.
\end{equation}
For the Rota--Baxter map $\tilde{R}$ we find $\tilde{R}(a *_R
b)=-\tilde{R}(a) \tilde{R}(b)$.


\subsection{Complete Rota--Baxter algebras and Spitzer's identity}

A {\bf{complete filtered Rota--Baxter algebra}}~\cite{EGK2} is
defined to be a Rota--Baxter algebra $(\mathcal{A},R)$ with a
complete decreasing filtration of Rota--Baxter ideals
$\{\mathcal{A}_n\}_{n\geq 0}$. So we have
$$
 \mathcal{A}_{n+1}\subset \mathcal{A}_n,\
 \mathcal{A}_m\mathcal{A}_n\subseteq \mathcal{A}_{m+n},\ R(\mathcal{A}_n)\subset
 \mathcal{A}_n,\ \lim_{\longleftarrow} \mathcal{A} / \mathcal{A}_n \cong \mathcal{A}.
$$
The last equation is equivalent to saying that $\mathcal{A}$ is
complete with respect to the topology on $\mathcal{A}$ defined by
the ideals $\mathcal{A}_n$.

\begin{exams} {\rm{Let $(A,R)$ be a
Rota--Baxter algebra with Rota--Baxter map $R$. We have the
following complete Rota--Baxter algebras.
\begin{enumerate}
\item The power series algebra $\mathcal{A}:=A[[x]]$ where the
filtration is given by the degree in $x$ and the Rota--Baxter
operator $\mathcal{R}:\mathcal{A} \to \mathcal{A}$ acts on a power
series via $R$ through the coefficients,
$\mathcal{R}\left(\sum_{n\ge 0}a_nx^n\right):=\sum_{n\ge
0}R(a_n)x^n$;
\item Let $\mathcal{H}_{\mathcal{F}}$ be the connected graded Hopf
algebra of Feynman graphs. The regularized functionals
$\Hom(\mathcal{H}_{\mathcal{F}},A)$ where the filtration is given
by the grading of $\mathcal{H}_{\mathcal{F}}$ and the Rota--Baxter
operator acts on a linear map $f:\mathcal{H}_{\mathcal{F}}\to A$
by acting on the target space image of $f$. See
Theorem~\ref{SpitzerpQFT};
\item The upper triangular matrices with entries in $A$ where the
filtration is given by the number of zero subdiagonals of the
matrices and the Rota--Baxter operator acts on a matrix entry by
entry. See \S~\ref{ss:matrixRB}.
\end{enumerate}
}}
\end{exams}

By the completeness of $(\mathcal{A},R)$, the functions
\begin{equation*}
 \exp: \mathcal{A}_1 \to 1+\mathcal{A}_1,\
 \exp(a):=\sum_{n=0}^\infty \frac{a^n}{n!},
\end{equation*}
\begin{equation*}
 \log: 1+\mathcal{A}_1 \to \mathcal{A}_1,\
 \log(1+a):=-\sum_{n=1}^\infty \frac{(-a)^n}{n}
\end{equation*}
are well-defined and are inverse of each other.

The algebraic formulation \cite{Baxter,Cartier,Kingman,Ro2} of the
{\bf{classical Spitzer identity}}~\cite{Sp} was first given for a
commutative Rota--Baxter algebra $A$. In terms of the complete
filtered commutative Rota--Baxter algebra $\mathcal{A}=A[[x]]$ in
the first example above, it takes the form
\begin{eqnarray}
 \exp\big(- \mathcal{R}(\log(1+ax)) \big)
         &=&\sum_{n=0}^\infty (-1)^n (Ra)^{[n]}x^n,\ \forall\, ax \in \mathcal{A}_1.
  \label{SpitzerId}
\end{eqnarray}
Here we inductively define
$$
  (R a)^{[n+1]}:=R\big((Ra)^{[n]}\:a\big)
$$
with the convention that $(Ra)^{[0]}=1$. We note that the element
$b:=\sum_{n=0}^\infty (-1)^n (Ra)^{[n]}x^n \in \mathcal{A}$
uniquely solves the recursive equation $b=1-\mathcal{R}(bax)$.

If $(\mathcal{A},R)$ is a complete filtered
commutative Rota--Baxter algebra of weight zero, i.e., $R$ fulfills
\begin{equation}
 \label{RB0}
 R(a)R(b)=R\big(R(a)b+aR(b)\big), \;\; \forall a,b \in \mathcal{A},
\end{equation}
then Spitzer's identity (\ref{SpitzerId}) reduces to the identity
$$
 \exp\big( - R(a)x \big)
    =\sum_{n=0}^\infty (-1)^n (Ra)^{[n]}x^n,\ \forall\, a \in \mathcal{A},
$$
which is well-known in the context of linear differential
equations, where $R$ is the Riemann integral. It follows naturally
from (\ref{RB0}) since $R(a)^n=n! (Ra)^{[n]}$ for $a\in A$.

In the following theorem, we generalize the Spitzer identity to
non-commutative complete filtered Rota--Baxter algebras. The
essential difference with the commutative case is the map
$\chi:\mathcal{A}_1 \to \mathcal{A}_1$ appearing inside the
exponential. This map was introduced in~\cite{EGK1} and is defined
recursively by
\begin{equation}
 \chi(u):=u - BCH\big(R(\chi(u)),\tilde{R}\,(\chi(u))\big), \;
 \forall u \in \mathcal{A}_1
 \label{BCHrecursion}
\end{equation}
using the {\bf{Baker--Campbell--Hausdorff}} (BCH) formula
$$
  \exp(x)\exp(y)=\exp\big(x+y+BC\!H(x,y)\big)
$$
which is a power series in $x,y \in \mathcal{A}_1$ of degree 2.
One finds a simpler recursion for $\chi$, using the factorization
property implied by the $\chi$ map on $\mathcal{A}$.
\begin{lema}\cite{EGK1} Let $\mathcal{A}$ be a complete filtered
$\mathbb{K}$-algebra. $K:\mathcal{A} \to \mathcal{A}$ is a linear
map. The map $\chi$ in (\ref{BCHrecursion}) solves the following
recursion
 \begin{equation}
   \label{BCHrecursion3}
   \chi(u):=u + BCH\big(-K(\chi(u)),u)\big),\;\; u\in \mathcal{A}_1.
 \end{equation}
\end{lema}
\begin{proof}
In general for any $u \in \mathcal{A}$  we can write $u = K(u) +
(\id_{\mathcal{A}}-K)(u)$ using linearity of $K$. The map $\chi$
then implies for $u \in \mathcal{A}_1$ that $\exp(u)=
\exp\big(K(\chi(u))\big) \exp\big(\tilde{K}(\chi(u))\big)$.
Further, \allowdisplaybreaks{
\begin{eqnarray*}
  \exp\big(\tilde{K}(\chi(u))\big)  &=&  \exp\big(-K(\chi(u))\big) \exp(u) \\
                                    &=& \exp\big(-K(\chi(u)) + u + BCH(-K(\chi(u)),u)\big).
\end{eqnarray*}}
Bijectivity of $\exp$ map then implies, that
 \allowdisplaybreaks{
\begin{eqnarray*}
 \chi(u) - K(\chi(u)) &=& -K(\chi(u)) + u + BCH\big(-K(\chi(u)),u\big).
\end{eqnarray*}}
From which Equation (\ref{BCHrecursion3}) follows.
\end{proof}
More details on this recursion can also be found in
\cite{EGK2,Manchon}. The factorization in the complete filtered
algebra $\mathcal{A}$ follows from the map $\chi$. Replacing the
linear map $K$ by a Rota--Baxter operator $R$, we arrive at
{\bf{Spitzer's identity for non-commutative Rota--Baxter
algebra}}.

{\begin{thm}
{\rm\cite{EGK2}} \label{complRB} Let
$(\mathcal{A},R,\mathcal{A}_n)$ be a complete filtered
Rota--Baxter algebra of weight one. Let $a \in \mathcal{A}_1$ and
recall that $\tilde{R}:=\id_\mathcal{A}-R$.
\begin{enumerate}
\item\label{eq:exp1} The equation $b=1-R(ba)$ has a unique
solution $b= \exp\big(-R(\chi(\log (1+a)))\big)$.
\item\label{eq:exp2} The equation $b'=1-\tilde{R}(ab')$ has a
unique solution $b'= \exp\big(-\tilde{R}(\chi(\log (1+a)))\big)$.
\end{enumerate}
\label{thm:equation} \label{spitzer}
\end{thm}}
When $(\mathcal{A},R)$ is commutative, the map $\chi$ reduces to
the identity map, giving back Spitzer's classical identity
(\ref{SpitzerId}). In general we have {\bf{Atkinson's
theorem}}~\cite{Atkinson}, giving a decomposition.
{\begin{thm} {\rm \cite{Atkinson,EGK2}}
\label{thm:Atkinson} Let $(\mathcal{A},R)$ be a complete
filtered
Rota--Baxter algebra. For solutions $b$ and $b'$ in items {\rm (\ref{eq:exp1})}
and {\rm (\ref{eq:exp2})} of Theorem \ref{complRB}, we have
\begin{equation}
  b (1+a) b' = 1, {\rm\ that\ is,\ } (1+a)=b^{-1}b'{}^{-1}.
\label{eq:atk2}
\end{equation}
If $R$ is idempotent: $R^2=R$, then this is the unique
decomposition of $1+a$ into a product of an element in
$1+R(\mathcal{A}_1)$ with an element in
$1+\tilde{R}(\mathcal{A}_1)$.
\end{thm}}

Recall that $\tilde{R}:=\id_\mathcal{A}-R$ is a Rota--Baxter
operator if and only if $R$ is. Further
$\tilde{\tilde{R}}:=\id_\mathcal{A}-\tilde{R} =R$. Thus by
exchanging $\tilde{R}$ and $R$ in the definition
(\ref{BCHrecursion}) of $\chi$ and in Theorem~\ref{thm:equation},
we have the following variation which will be useful later.
{\begin{thm} \label{complRBo} Let $(\mathcal{A},R,\mathcal{A}_n)$
be a complete filtered Rota--Baxter algebra of weight one. Define
$\bar{\chi}:\mathcal{A}_1\to \mathcal{A}_1$ by the recursion
\begin{equation}
 \bar{\chi}(u):=u - BCH\big(\tilde{R}(\bar{\chi}(u)),R(\bar{\chi}(u))\big), \;
 \forall u \in \mathcal{A}_1.
 \label{BCHrecursiono}
\end{equation}
Let $a \in \mathcal{A}_1$.
\begin{enumerate}
\item\label{eq:exp1o} The equation $\bar{b}=1-R(a\bar{b})$ has a
                      unique solution $\bar{b}= \exp\big(-R(\bar{\chi}(\log (1+a)))\big)$.
\item\label{eq:exp2o} The equation
                      $\bar{b}'=1-\tilde{R}(\bar{b}'a)$ has a unique solution
                      $\bar{b}'=\exp\big(-\tilde{R}(\bar{\chi}(\log (1+a)))\big).$
\item\label{Atkinson} For solutions $\bar{b}$ and $\bar{b}'$ in
                      item {\rm (\ref{eq:exp1o})} and {\rm (\ref{eq:exp2o})}, we have
\begin{equation}
  \bar{b}' (1+a) \bar{b} = 1, {\rm\ that\ is,\ }
    (1+a)=\bar{b}'{}^{-1}\bar{b}^{-1}.
\label{eq:atk2o}
\end{equation}
When $R$ is idempotent, this gives the unique decomposition of
$1+a$ into a product of an element in $1+\tilde{R}(\mathcal{A}_1)$
with an element in $1+R(\mathcal{A}_1)$.
\end{enumerate}
\label{thm:equationo} \label{spitzero}
\end{thm}}

The two decompositions in Theorem~\ref{thm:Atkinson} and
Theorem~\ref{thm:equationo} are simply related as follows.
\begin{prop} \label{opposite}
Let $\mathcal{A}^{\rm op}$ be the opposite algebra of
$\mathcal{A}$, with product defined by $a\centerdot b:=ba$. Let
$O:\mathcal{A}\to \mathcal{A}^{\rm op}, O(a)=a$, be the canonical
antihomomorphism of Rota--Baxter algebras. For $a\in
\mathcal{A}_1$, let $1+a=b^{-1}b'{}^{-1}$ be the decomposition in
Theorem~\ref{thm:Atkinson}, with $b$ (resp. $b'$) being solution
of item (1) (resp. (2)) of Theorem~\ref{thm:equation}. Then in
$$
  1+a=O(1+a)= O\big(b'{}^{-1}\big)\centerdot O\big(b^{-1}\big)=O(b')^{-1}\centerdot O(b)^{-1},
$$
the factor $O(b)$ (resp. $O(b')$) is the solution, in the opposite
Rota--Baxter algebra $(\mathcal{A}^{\rm op},R)$, from item (2)
(resp. item (1)) of Theorem~\ref{thm:equationo}.
\label{pp:relation}
\end{prop}
\begin{proof}
We just need to note that, under the anti-isomorphism
$O:\mathcal{A}\to \mathcal{A}^{\rm op}$, the defining equations of
$\chi$, $b$ and $b'$ in Theorem~\ref{thm:equation}, with
multiplication in $\mathcal{A}$, are sent to the defining
equations of $\bar{\chi}$, $\bar{b}$ and $\bar{b}'$ in
Theorem~\ref{thm:equationo}, with multiplication in
$\mathcal{A}^{\rm op}$.
\end{proof}
We give another variation of Theorem~\ref{complRB}.
{\begin{prop} \label{inverse} Let $b$ and $b'$ be solutions of the
equations in item {\rm (\ref{eq:exp1})} respectively {\rm
(\ref{eq:exp2})} of Theorem \ref{complRB}. Let
$\check{a}=(1+a)^{-1}-1$. Then
 \begin{enumerate}
  \item $b^{-1}= \exp\big(R(\chi(\log (1+a)))\big)$ is the unique
        solution of the equation $c=1-R\big(\check{a}c\big)$.
  \item $b'{}^{-1}= \exp\big(\tilde{R}(\chi(\log (1+a)))\big)$
        is the unique solution of the equation $c' =1-\tilde{R}\big(c'\check{a}\big)$.
  \item Further
$$
  b^{-1} = 1+R\big(a\ b'), \qquad b'{}^{-1} = 1+\tilde{R}\big(b\ a).
$$
 \end{enumerate}
\end{prop}}
\begin{proof}
(1) Since $\check{a}$ is in $\mathcal{A}_1$, the equation
$c=1-R(\check{a}c)$ has a unique solution. We just need to check
that the solution $c$ is the inverse of $b$. Since $b$ satisfies
$b=1-R(ba)$, we have \allowdisplaybreaks{
\begin{eqnarray*}
  bc &=& (1-R(ba))(1-R(\check{a}c))\\
     &=& 1-R(ba)-R(\check{a}c)+R(ba)R(\check{a}c)\\
     &=& 1-R(ba)-R(\check{a}c)+R(baR(\check{a}c))+R(R(ba)\check{a}c)-R(ba\check{a}c)\\
     &=& 1-R\big(ba(1-R(\check{a}c)\big) - R\big((1-R(ba))\check{a}c\big) - R(ba\check{a}c) \\
     &=& 1-R(bac)-R(b\check{a}c)-R(ba\check{a}c) \\
     &=& 1-R\big(b(a+\check{a}+a\check{a})c\big),
\end{eqnarray*}}
since $a+\check{a}+a\check{a}=a+(1+a)\check{a}=a-a=0$, we get
$bc=1$, as needed.

(2)
The prove of the second statement is the same.

(3) Note that
$$
 \check{a}b^{-1}= (b'b-1)b^{-1}=b'-b^{-1} =b'-(1+a)b'=-ab'.
$$
So by item (1), we have
$$
  b^{-1}=1-R(\check{a}b^{-1})= 1+R(ab').
$$
\end{proof}

By Theorem~\ref{thm:equationo}.(\ref{eq:exp1o}), the equation
$c=1-R(\check{a}c)$ in Proposition~\ref{inverse} has a unique solution
$$
  c=\exp\big(-R(\bar{\chi}(\log(1+\check{a})))\big)
   = \exp\big(-R(\bar{\chi}(\log(1+a)^{-1}))\big)
   = \exp\big(-R(\bar{\chi}(-\log(1+a)))\big).
$$
By the bijectivity of $\log$ and $\exp$, we have
$$
  R\big(\chi(u)\big)=-R\big(\bar{\chi}(-u)\big), \ \forall\, u\in \mathcal{A}_1.
$$
Thus we obtain
\begin{corl}
$$
  R\big(\chi(u)+\bar{\chi}(-u)\big)=0,\ \forall\, u\in \mathcal{A}_1.
$$
\end{corl}


\subsection{Decomposition of triangular matrices}\label{ss:matrixRB}

Let $A$ be a commutative $\mathbb{K}$-algebra. In the following
one might replace ''upper" by ''lower" without restriction. The
algebra of $n \times n$ matrices with entries in $A$ is denoted by
$\mathcal{M}_n(A)$. We have the subalgebras $\mathcal{M}^u_n(A)
\subset \mathcal{M}_n(A)$, $1\leq n < \infty$, of upper triangular
matrices. We also let $\mathcal{M}^u_\infty(A)$ denote the algebra
of $\infty \times \infty$ upper triangular matrices.

The subset $\mathfrak{M}_n(A)$, $1\leq n\leq \infty$, of upper
triangular matrices with unit diagonals, i.e. those $\alpha \in
\mathcal{M}^u_n(A)$ such that $\alpha_{ii}=1$, $i=1,\dots,n$, form
a group under matrix multiplication. The inverse of
$\alpha=(\alpha_{ij}) \in \mathfrak{M}_n(A)$ is given by the
well-known, recursively defined inversion formula for upper
triangular $n \times n$ matrices
\begin{equation}
  \label{antipodeInverse}
  (\alpha^{-1})_{ij}=-\alpha_{ij}-\sum_{k=i+1}^{j-1}(\alpha^{-1})_{ik}\alpha_{kj}.
\end{equation}
Here commutativity of the algebra $A$ is needed. \smallskip

For each $n\leq \infty$, the algebra $\mathcal{M}^u_n(A)$ carries
a natural decreasing filtration in terms of the number of zero
upper subdiagonals. We denote by $\mathcal{M}^u_n(A)_1$ the upper
triangular matrices with the main diagonal being zero, that is,
the strict upper triangular matrices. Let $\mathcal{M}^u_n(A)_k$,
$k>1$, denote the ideal of strictly upper triangular matrices with
zero on the main diagonal and on the first $k-1$ subdiagonals. We
then have the decreasing filtration
$$
   \mathcal{M}^u_n(A) \supset \mathcal{M}^u_n(A)_1 \supset \dots
  \supset \mathcal{M}^u_n(A)_{k-1} \supset \mathcal{M}^u_n(A)_k \supset \cdots, k< n,
$$
with
$$
  \mathcal{M}^u_n(A)_k\: \mathcal{M}^u_n(A)_m \subset
  \mathcal{M}^u_n(A)_{k+m}.
$$
We also have $\mathfrak{M}_n(A) = {\bf 1} + \mathcal{M}^u_n(A)_1$,
here ${\bf 1}$ denotes the $n\times n$ unit matrix. It is easy to
see that for any $n\leq \infty$, the filtration is complete, that
is, $\mathcal{M}^u_n(A)$ is complete with respect to the topology
defined by the ideals $\mathcal{M}^u_n(A)_k,\, k\geq 0$.

Thus the maps
\begin{equation}
  \label{expDef}
    \exp: \mathcal{M}^u_n(A)_1 \to \mathfrak{M}_n(A),\
    \exp(Z)=\sum_{k=0}^\infty \frac{Z^n}{n!},
\end{equation}
\begin{equation}
  \label{logDef}
    \log: \mathfrak{M}_n(A) \to \mathcal{M}^u_n(A)_1,\
    \log(\alpha)= - \sum_{k=1}^\infty \frac{(1-\alpha)^n}{n}.
\end{equation}
are well-defined and are the inverse of each other. We denote
$Z_\alpha=\log(\alpha)$ for $\alpha\in \mathfrak{M}_n(A)$.

Now let $A$ be a Rota--Baxter algebra with Rota--Baxter operator
$R$. We define a Rota--Baxter map $\mathcal{R}$ on
$\mathcal{M}^u_n(A)$ by extending the Rota--Baxter map $R$
entrywise, i.e. for the matrix $\alpha =(\alpha_{ij}) \in
\mathcal{M}^u_n(A)$, define
\begin{equation}
  \label{RBonMatrices}
 \mathcal{R}(\alpha) =  \big ( R(\alpha_{ij})\big).
\end{equation}
\begin{thm} \label{thm:RBalgMatrices}
 The triple $\left(\mathcal{M}^u_n(A),\mathcal{R},\{\mathcal{M}^u_n(A)_k\}_{k\geq 1}\right)$
 forms a complete filtered Rota--Baxter algebra.
\end{thm}
\begin{proof}
We only need to show the Rota--Baxter relation for $\mathcal{R}$.
For distinction, we use brackets $[$ and $]$ for matrix
delimiters. Let $\alpha=[\alpha_{ij}]$ and $\beta=[\beta_{ij}]$ be
in $\mathcal{M}^u_n(A)$. By the entry-wise definition of
$\mathcal{R}$ in (\ref{RBonMatrices}) and the Rota--Baxter
relation, we have \allowdisplaybreaks{
\begin{align*}
\mathcal{R}(\alpha)\:\mathcal{R}(\beta) &=\big [R(\alpha_{ij})\big] \big[R(\beta_{ij})\big] \\
&= \Big [\sum_{k} R(\alpha_{ik})R(\beta_{kj})\Big] \\
&= \Big [ \sum_{k} \big( R(R(\alpha_{ik})\beta_{kj})
    + R(\alpha_{ik}R(\beta_{kj}))-R(\alpha_{ik}\beta_{kj}) \big)\Big ]\\
&= \Big [\sum_k R(R(\alpha_{ik})\beta_{kj})\Big]
    + \Big [\sum_k R(\alpha_{ik}R(\beta_{kj}))\Big ]
    - \Big [\sum_k  R(\alpha_{ik}\beta_{kj})\Big ]\\
&= \Big [R\big( \sum_k R(\alpha_{ik})\beta_{kj} \big) \Big ]
    + \Big [R\big( \sum_k \alpha_{ik}R(\beta_{kj}) \big)\Big ]
    - \Big [ R\big( \sum_k \alpha_{ik}\beta_{kj} \big)\Big ]\\
&= \mathcal{R} \Big [\sum_k R(\alpha_{ik})\beta_{kj} \Big ]
    + \mathcal{R} \Big [\sum_k \alpha_{ik}R(\beta_{kj})\Big ]
    - \mathcal{R} \Big [ \sum_k \alpha_{ik}\beta_{kj}\Big ]\\
&= \mathcal{R} \big( \mathcal{R} ([\alpha_{ij}])[\beta_{ij}]\big)
    + \mathcal{R} \big( [ \alpha_{ij}]\mathcal{R}([\beta_{ij}])\big)
    - \mathcal{R} \big( \big[\alpha_{ij}\big]\big[\beta_{ij} \big] \big)\\
&=\mathcal{R}(\mathcal{R}(\alpha)\beta)+
      \mathcal{R}(\alpha\mathcal{R}(\beta))-\mathcal{R}(\alpha\,\beta)
\end{align*}}
as needed.
\end{proof}
Then we can apply Theorems \ref{complRB}, \ref{thm:Atkinson} and
\ref{complRBo} to obtain decompositions of upper triangular
matrices. For later applications to matrix Birkhoff decomposition
in renormalization, we will stress the variation in
Theorem~\ref{complRBo}.

\begin{corl}
Let $\alpha$ be in $\mathfrak{M}_n(A)$.
\begin{enumerate}
\item There is a factorization
\begin{equation}
 \label{two}
 \alpha = \bar{\alpha}_+\bar{\alpha}_{-}^{-1},
\end{equation}
of $\alpha$ into a product of an element in
${\bf{1}}+\tilde{\mathcal{R}}\big(\mathcal{M}^u_n(A)_1\big)$ and
an element in ${\bf{1}} +
\mathcal{R}\big(\mathcal{M}^u_n(A)_1\big)$ which is unique if
$R^2=R$. \item The factors $\bar{\alpha}_+$ and $\bar{\alpha}_-$
have the explicit expression
\begin{equation}
  \bar{\alpha}_+=\exp\Big(\tilde{\mathcal{R}}\big(\bar{\chi}(Z_\alpha)\big)\Big), \quad
  \bar{\alpha}_-=\exp\Big(-\mathcal{R}\big(\bar{\chi}(Z_\alpha)\big)\Big).
\label{expFormulas}
\end{equation}
Here $Z_\alpha=\log(\alpha)$ and $\bar{\chi}$ is defined in
Equation~(\ref{BCHrecursiono}) in analogy to $\chi$. \item
Further, $\bar{\alpha}_+^{-1}$ is the unique solution to the
equation
\begin{equation}
\bar{\beta}'={\bf 1}-\tilde{\mathcal{R}}(\bar{\beta}'(\alpha-{\bf
1})) \label{eq:barplus}
\end{equation}
and $\bar{\alpha}_-$ is the unique solution to the equation
\begin{equation}
\bar{\beta}={\bf 1}-\mathcal{R}((\alpha-{\bf 1})\bar{\beta}).
\label{eq:barminus}
\end{equation}
\end{enumerate}
\label{matdecomp}
\end{corl}
Equations (\ref{eq:barplus}) and (\ref{eq:barminus}) are similar
to the well-known recursions in renormalization theory where they
are called Bogoliubov recursions \cite{Collins,CKI}. We will
explore their connection in the next section through a matrix
representation of regularized Feynman characters in
renormalization in perturbative QFT.

Corollary~\ref{matdecomp} suggests that $\bar{\alpha}_-$ and
$\bar{\alpha}_+$ in (\ref{two}) can be calculated either by their
exponential formulae (\ref{expFormulas}) or directly from their
recursive equations (\ref{eq:barplus}) and (\ref{eq:barminus}). We
will first describe the recursive method to find the factor
matrices $\bar{\alpha}_-$ and $\bar{\alpha}_+$. Later in
subsection \ref{normal} we will calculate the factor matrices
using the BCH-recursion $\bar{\chi}$ in
(\ref{BCHrecursiono}).\medskip

As an example we first consider a straightforward $2 \times 2$
factorization of the matrix $\alpha  \in \mathfrak{M}_2(A)$,
$$
 \alpha=\left(
    \begin{array}{cc}
      1&a\\
      0&1
    \end{array}
        \right) =
    \left(
     \begin{array}{cc}
       1&R(a)\\
       0&1
     \end{array}
    \right)\
    \left(
    \begin{array}{cc}
      1&\tilde{R}(a)\\
      0&1
    \end{array}
        \right),
$$ which simply follows from $R+\tilde{R}=\id_A$.

The case of $3\times 3$ matrices is already more telling. For a
given $\alpha$~{\small{$= \left(
    \begin{array}{ccc}
      1&a&b\\
      0&1&c\\
      0&0&1
    \end{array}
        \right)$}} in $\mathfrak{M}_3(A)$, the
equation that
 $\bar{\alpha}_-$~{\small{$:=\left (
        \begin{array}{ccc}
           1&\bar{a}&\bar{b}\\
           0&1 &\bar{c} \\
           0&0 &1
        \end{array} \right)$}} should fulfill is (\ref{eq:barminus}):
$$
  \left (
   \begin{array}{ccc}
       1 & \bar{a} & \bar{b} \\
       0 & 1  & \bar{c}  \\
       0 & 0  & 1
   \end{array} \right )
      = {\bf{1}} - \mathcal{R}\left(
      \left(
         \begin{array}{ccc}
         0 & a  & b\\
         0 & 0  & c\\
         0 & 0  & 0
      \end{array} \right)
      \left(
         \begin{array}{ccc}
         1 & \bar{a} & \bar{b}\\
         0 & 1  & \bar{c} \\
         0 & 0  & 1
      \end{array} \right)
  \right).
$$
Equating the two sides entry-wise, we simply obtain,
$$
  \bar{a}=-R(a),\; \bar{c}=-R(c),\; \bar{b}=-R(a\bar{c}+b)=R(aR(c))-R(b).
$$
So
$$ \bar{\alpha}_-=      \left(
         \begin{array}{ccc}
         1 & -R(a) & -R(b)+R(aR(c))\\
         0 & 1  & -R(c) \\
         0 & 0  & 1
      \end{array} \right)
$$
and by using Equation (\ref{antipodeInverse}), we obtain
$$ \bar{\alpha}^{-1}_-=
  \left(
        \begin{array}{ccc}
         1 & R(a) & R(b) +  R(aR(c)) -R(a)R(c)\\
         0 & 1 & R(c)\\
         0 & 0 & 1
        \end{array}
  \right)
=
  \left(
        \begin{array}{ccc}
         1 & R(a) & R(b) +  R(R(a)c) -R(ac)\\
         0 & 1 & R(c)\\
         0 & 0 & 1
        \end{array}
  \right)
$$
We similarly use (\ref{eq:barplus}) to find
$$\bar{\alpha}^{-1}_+=
\left(
         \begin{array}{ccc}
         1 & -\tilde{R}(a) & -\tilde{R}(b)+\tilde{R}(\tilde{R}(a)c)\\
         0 & 1  & -\tilde{R}(c) \\
         0 & 0  & 1
      \end{array} \right)
$$
and then use (\ref{antipodeInverse}) to find
$$
 \bar{\alpha}_+=
    \left(
         \begin{array}{ccc}
         1 & \tilde{R}(a) & \tilde{R}(b)+\tilde{R}(a\tilde{R}(c))-\tilde{R}(ac)\\
         0 & 1  & \tilde{R}(c) \\
         0 & 0  & 1
         \end{array} \right)
$$
We thus obtain the unique factorization in (\ref{two}):
\begin{equation}
\alpha=\bar{\alpha}_+\bar{\alpha}^{-1}_-
=\left(
         \begin{array}{ccc}
         1 & \tilde{R}(a) & \tilde{R}(b)+\tilde{R}(a\tilde{R}(c))-\tilde{R}(ac)\\
         0 & 1  & \tilde{R}(c) \\
         0 & 0  & 1
      \end{array} \right)
  \left(
        \begin{array}{ccc}
         1 & R(a) & R(b) +  R(R(a)c) -R(ac)\\
         0 & 1 & R(c)\\
         0 & 0 & 1
        \end{array}
  \right)
\label{eq:barminuse}
\end{equation}
To compute $\bar{\alpha}_\pm$ effectively in general, we have
\begin{thm} Let $\alpha \in \mathfrak{M}_n(A)$.
\allowdisplaybreaks{
\begin{enumerate}
   \item The equation
   $\bar{\beta}={\bf{1}}-\mathcal{R}\big((\alpha-{\bf{1}}) \: \bar{\beta} \big)$
   has a unique solution $\bar{\beta}=(\bar{\beta}_{ij})$ where
  \begin{equation}
    \label{barRec1}
    \bar{\beta}_{ij}=  -R(\alpha_{ij}) - \sum_{k=2}^{j-i}\: \sum_{i<l_1<l_2< \cdots <l_{k-1}<j} (-1)^{k+1}
    R\big(\alpha_{il_1}R(\alpha_{l_1l_2}R(\alpha_{l_2l_3} \cdots R(\alpha_{l_{k-1}j})\cdots ))\big).
  \end{equation}

 \item The equation $\bar{\beta}'={\bf{1}}-\tilde{\mathcal{R}}\big(\bar{\beta}' \:
                                                                          (\alpha-{\bf{1}})\big)$
         has a unique solution $ \bar{\beta}'=(\bar{\beta}'_{ij})$ where
  \begin{equation}
     \label{barRec2}
       \bar{\beta}'_{ij}= -\tilde{R}(\alpha_{ij})
                                - \sum_{k=2}^{j-i}\: \sum_{i<l_1<l_2<\cdots<l_{k-1}<j} (-1)^{k+1}
       \tilde{R}\big(\tilde{R}(\cdots \tilde{R}(\tilde{R}(\alpha_{i l_{1}})\alpha_{l_{1} l_{2}})
                                              \cdots \alpha_{l_{k-2} l_{k-1}})\alpha_{ l_{k-1} j}\big).
  \end{equation}
\end{enumerate}}
\label{thm:matrix}
\end{thm}
\begin{proof}
We will prove Equation (\ref{barRec1}). The proof for the second
equation is the same. Comparing two sides of $ \bar{\beta}={\bf
1}-\mathcal{R}((\alpha-{\bf 1})\bar{\beta})$, we have
$$\bar{\beta}_{ij}=\left \{ \begin{array}{ll}
    0,& i>j,\\
    1,& i=j,\\
    -R\Big(\alpha_{ij}+\sum_{i<u<j}\alpha_{iu}\bar{\beta}_{uj}\Big), & i<j
    \end{array} \right .
$$
So we just need to prove Equation (\ref{barRec1}) for $i<j$. For
this we use induction on $j-i\geq 1$. There is nothing to prove
when $j-i=1$. Assuming the equation holds for $j-i\leq m$, then
for $j-i=m+1$, we have
\begin{eqnarray*}
\bar{\beta}_{ij}\!\!\!&=&\!\!\!\! -R\Big(\alpha_{ij}+\sum_{i<u<j}\alpha_{iu}\bar{\beta}_{uj}\Big) \\
                \!\!\!&=&\!\!\!\! -R(\alpha_{ij})-R\bigg(\!\sum_{i<u<j}\!\! \alpha_{iu}\Big(\!\!
   -R(\alpha_{uj}) \!-\! \sum_{k=2}^{u-i}\: \sum_{u<l_1<l_2< \cdots <l_{k-1}<j}\!\!\!\!\!
    (-1)^{k+1}  R\big(\alpha_{ul_1}R(\alpha_{l_1l_2}
                                    \cdots R(\alpha_{l_{k-1}j})\cdots \!)\big)\!\Big)\!\!\bigg).
\end{eqnarray*}
This is what we need.
\end{proof}
The same formulae hold when $R$ and
$\tilde{R}$ are exchanged. These formulae will be useful in the
next section when we consider anti-homomorphisms of Rota--Baxter
algebras.
Let us also remark at this stage that the simplicity of the
formulae in Theorem \ref{thm:matrix} suggests a computational
advantage of working with matrices over working directly with regularized
Feynman characters.


\section{Matrix calculus in perturbative QFT}\label{CK-BC}

The perturbative approach to quantum field theory provides
theoretical predictions which match the experimental data with
very high precision. At the heart of this scheme lies the idea to
approximate physical quantities by power series expansions in
terms of a supposed to be small parameter called the coupling
constant, which measures the strength of interaction in the
physical system under investigation. Of course, this idea raises
immediately the question of convergence of such power series, but
for such issues we refer the interested reader to more
sophisticated presentations of perturbative QFT. The power series
expansion is organized by Feynman graphs which serve to labels
the terms in the power series. A Feynman graph is a collection of
vertices and edges, reflecting the interaction and propagation of
particles, respectively. For each term in the power series the
number of loops of the corresponding Feynman graph fixes the order
in the power series. The terms of the power series, called Feynman
amplitudes, follow from Feynman rules which provide a way to
translate a Feynman graph into a Feynman integral.

In general this perturbative ansatz is plagued with so-called
ultraviolet divergencies, i.e. the Feynman integrals corresponding
to Feynman graphs diverge in the limit of large momenta or
equivalently small distances, and therefore seem to be useless in
physics. For now we ignore the infrared problems
appearing in theories with massless particles.

Renormalization theory in perturbative QFT was developed to cure
these divergencies in a meaningful way. Actually, it consists of
two steps, first a regularization prescription meant to control
the divergences by extending the target space of the Feynman rules
from the base field $\mathbb{C}$ to a particular algebra of
regularized Feynman amplitudes. As a main example we mention here
dimensional regularization, where this algebra is that of the
Laurent series. Second, on the new target space algebra, a
renormalization scheme is introduced, that isolates the
problematic pieces, i.e., the divergent part. The renormalization
process itself is made up in a recursive manner based on the self
similar structure of Feynman graphs. Hereby we mean the fact that
graphs of lower order --in terms of the number of loops-- appear
inside Feynman graphs of higher order.

Unfortunately, despite its impressive successes, renormalization
was stigmatized, especially for its lack of a sound mathematical
underpinning. One reason for this weakness might have been the
fact that Feynman graphs in itself appeared to be unrelated to any
mathematical structure that may possibly underlie the
renormalization prescription.

This changed to a great extend with the original paper by Kreimer
\cite{Kreimer1}, followed by the work of Connes and Kreimer
\cite{CKI,CKII,CKIII,CK4}. The combinatorial-algebraic side of the
process for renormalization is captured via a combinatorial Hopf
algebra of Feynman graphs, essentially characterized by the
coproduct map which organizes the decomposition of a Feynman graph
into its subgraphs in a firm way. The analytical side of
renormalization, i.e. the Feynman rules providing the Feynman
integrals, is imbedded as algebra homomorphisms in the dual space
of this Feynman graph Hopf algebra.


\subsection{Connes--Kreimer renormalization Hopf algebra}

Let us briefly summarize the main results of the Hopf algebra
description of renormalization theory. For a general introduction
to Hopf algebras the reader might want to consult the standard
references like for instance \cite{Abe,FGV,Sw}. Details about Hopf
algebras in renormalization theory can be found in the survey
articles \cite{FG,Manchon}.


\subsubsection{Feynman graphs and decorated rooted
trees}\label{FeynTrees}

Let us mention for completeness how Feynman graphs and decorated
rooted trees are related. We assume that the reader is somewhat
familiar with the former. The latter will be introduced in the
next section. For more details on Feynman graphs, such as the
notion of ultraviolet (UV) subgraphs, we refer to the standard
literature such as \cite{Collins}. Kreimer~\cite{Kreimer3}, and
then Connes--Kreimer~\cite{CKI,CKII} well-developed this link.

A Feynman graph is a collection of internal and external lines, or
edges, and vertices of several types. A proper subgraph of a graph
is determined by proper subsets of the set of internal edges and
vertices. Of vital importance are so-called one particle
irreducible (1PI) Feynman graphs, a connected graph that cannot be
made disconnected by removing one of its internal edges. In
general, a Feynman graph $\Gamma$ beyond one-loop order, is
characterized essentially by the appearance of its UV Feynman
subgraphs $\gamma_i \subset \Gamma$. For instance, two proper
Feynman subgraphs $\gamma_1,\ \gamma_2 \subset \Gamma$ might be
nested, $\gamma_1 \subset \gamma_2 \subset \Gamma$, or disjoint,
$\gamma_1 \cap \gamma_2 =\emptyset$. This hierarchy in which
subgraphs appear inside another subgraph is best represented by a
decorated rooted tree. The example below is borrowed from
$\varphi_{4dim}^4$-theory.
\begin{figure}[!h]
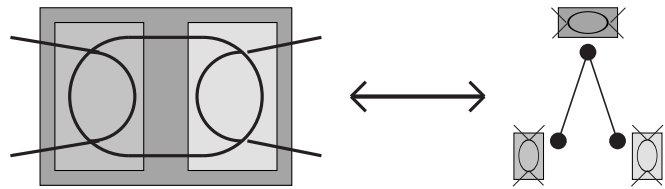

\begin{center}
     \scalebox{2}{\aaaaa}
\end{center}
\caption{{\small{Example of a 3-loop graph with two nested
disjoint UV subgraphs from $\varphi_{4dim}^4$-theory.}}}
\end{figure}

\noindent The two UV subgraphs $\phantom{.}$ \Anewfish (little
boxes inside) are disjoint, but nested inside another Feynman
graph of the same type, \newfish. The rooted tree on the right
represents this hierarchy. There is a third possibility, that
subgraphs might be overlapping. Such Feynman graphs are
represented by linear combinations of decorated rooted trees. Let
us illustrate this with an example from $\varphi_{6dim}^3$-theory.
\begin{figure}[!h]
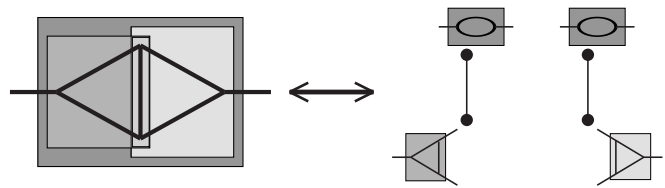

\begin{center}
     \scalebox{2}{\fishfishtree3}
\end{center}
\caption{{\small{2-loop example with overlapping UV subgraphs from
$\varphi_{6dim}^3$-theory.}}}
\end{figure}

\noindent For a detailed treatment of this special and
important case of overlapping structures we refer to~\cite{Kreimer3}.\\
The essential combinatorial operation on the set $\mathcal{F}$ of
(equivalence classes of) Feynman graphs in the process of
renormalization is a particular decomposition of such a Feynman
graph into its UV subgraphs, well-known to the practitioners under
the name of Bogoliubov's $\bar{\mathrm{R}}$-operation, or its
solution by Zimmermann's forrest formula~\cite{Zimmermann}. The
concept of the combinatorial Hopf algebra of Feynman graphs,
$\mathcal{H}_\mathcal{F}$, enters via its coproduct map, denoted
$\Delta:\mathcal{H}_\mathcal{F} \to \mathcal{H}_\mathcal{F}\otimes
\mathcal{H}_\mathcal{F}$, which organizes such a decomposition in
a mathematical sound way, e.g. for the last examples from
$\varphi_{6dim}^3$-theory we find
\delete{ \allowdisplaybreaks{
\begin{eqnarray*}
 \Delta\left(\!\!\!\begin{array}{c}\acoprod \\ \end{array}\!\!\! \right)
 \!\!\!&=&\!\!\!\!
\begin{array}{c}\acoprod \\ \end{array}\!\! \otimes 1_{\mathcal{F}}
  +
  1_{\mathcal{F}} \otimes \!\!\begin{array}{c}\acoprod \\ \end{array}
  \!+\!\!\!
\begin{array}{c}\lcop1 \\ \end{array}\!\! \otimes
\!\!\!\!\!\!\begin{array}{c}\rr1cop \\ \end{array}
  \!+\!\!\!
\begin{array}{c}\lcop1 \\ \end{array}\!\! \otimes
\!\!\!\!\!\!\begin{array}{c}\rl1cop \\ \end{array}
 \!+\!\!\!
\begin{array}{c}\lcop1 \lcop1\\ \end{array}\!\! \otimes
\!\!\begin{array}{c}\rcop1 \\ \end{array}
\end{eqnarray*}}}

\allowdisplaybreaks{
\begin{eqnarray*}
 \Delta\left(\!\!\!\begin{array}{c}\coprod2 \\ \end{array}\!\!\! \right)
 \!\!\!&=&\!\!\!\!
\begin{array}{c}\coprod2 \\ \end{array}\!\! \otimes 1_{\mathcal{F}}
  +
  1_{\mathcal{F}} \otimes \!\!\begin{array}{c}\coprod2 \\ \end{array}
  \!+\!
\begin{array}{c}\l2cop \\ \end{array}\!\! \otimes
\!\!\begin{array}{c}\c2cop \\ \end{array}
  \!+\!
\begin{array}{c}\r2cop\\ \end{array}\!\! \otimes
\!\!\begin{array}{c}\c2cop \\ \end{array},
\end{eqnarray*}}
\noindent which must be compared with Bogoliubov's formula for the
counter term $C(\Gamma)$ of a Feynman graph $\Gamma$. Only here we
apply this notation for the counter term used in \cite{CKI}, and
we use a symbolic graph notation. $R$ denotes the renormalization
scheme map
\allowdisplaybreaks{
\begin{eqnarray*}
 C\left(\!\!\!\begin{array}{c}\coprod2 \\ \end{array}\!\!\! \right)
 \!\!\!&=&\!\!\!\!
-R\left(\!\!\!\begin{array}{c}\coprod2 \\ \end{array}\!\!\!\right)
+R\left(R\left(\!\!\!\begin{array}{c}\l2cop \\
\end{array}\!\!\right)\!\!\begin{array}{c}\c2cop \\
\end{array}\right)
+R\left(R\left(\!\!\begin{array}{c}\r2cop\\ \end{array}\!\!
\right) \!\!\begin{array}{c}\c2cop \\ \end{array}\right)\\[-0.3cm]
\end{eqnarray*}}
Kreimer~\cite{Kreimer1} was the first to realize in
renormalization this underlying Hopf algebra structure.

From a conceptual point of view, and having the non-specialist in
mind, we feel that it is useful to work with decorated rooted
trees. Also, this underlines the generality of our results,
applicable to any theory in perturbative QFT and its
renormalization by choosing the particular set of decorations
dictated by the theory itself. Nevertheless, in terms of
applications of Connes--Kreimer's Hopf algebra of renormalization
to physics it is most naturally to formulate the Hopf algebra
directly on Feynman graphs, once a theory has been specified. We
included a simple calculation in section \ref{FeymanExample} using
an example from $\varphi^4_{4dim}$-theory. Also, we refer the
reader to the companion paper~\cite{Echo} with
J.~M.~Gracia-Bond\'ia and J.~C.~V\'arilly, which contains detailed
applications of the presented results to renormalization in
perturbative QFT. We should warn the reader, that in~\cite{Echo} a
slightly different notation is used.


\subsubsection{Connes--Kreimer Hopf algebra of rooted trees}

A rooted tree $t$ is made out of vertices and nonintersecting
oriented edge, such that all but one vertex have exactly one
incoming edge. We denote the set of vertices and edges of a rooted
tree by $V(t)$, $E(t)$ respectively. The root is the only vertex
with no incoming line. The empty tree is denoted by
$1_{\mathcal{T}}$. Each rooted tree is a representant of an
isomorphism class, and the $\mathbb{K}$-vector space freely
generated by the set of all isomorphism classes will be denoted by
$\mathcal{T}$.
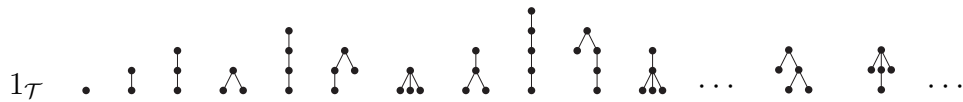
$$
 1_{\mathcal{T}}\;\;\;\;
 \ta1           \;\;\;\;
 \tb2           \;\;\;\;
 \tc3           \;\;\;\;
 \td31          \;\;\;\;
 \te4           \;\;\;\;
 \tf41          \;\;\;\;
 \th43          \;\;\;\;
 \thj44          \;\;\;\;
 \ti5           \;\;\;\;
 \tj51         \;\;\;\;
 \tm54         \;\;\; \cdots \;\;\; \tp56  \;\;\; \tr58 \;\;\; \cdots
$$
\begin{defn}
The $\mathbb{K}$-algebra of
non-planar rooted trees, denoted by $\mathcal{H}_{\mathcal{T}}$, is
the polynomial algebra, generated by the symbols $t$, each
representing an isomorphism class in $\mathcal{T}$. The unit is the
empty tree $1_{\mathcal{T}}$. The disjoint union of
rooted trees serves as a product, denoted by juxtaposition, i.e.
$m_{\mathcal{A}_{\mathcal{T}}}(t',t)=:t't$.
\end{defn}
There exists a natural grading on $\mathcal{H}_{\mathcal{T}}$ in
terms of the number of vertices of a rooted tree, $\#(t):=|V(t)|$.
On forests of rooted trees, we extend it to $\#(t_1 \cdots
t_n):=\sum_{i=1}^{n}\#(t_i)$. So that
$\mathcal{H}_{\mathcal{T}}=\mathbb{K} \oplus \bigoplus_{n
> 0} \mathcal{H}^{(n)}_{\mathcal{T}}$ becomes a graded
commutative $\mathbb{K}$-algebra.

To define a coproduct on $\mathcal{H}_\mathcal{T}$, we introduce
the notion of admissible cuts on a rooted tree. A cut $c_t$ on a
rooted tree $t \in \mathcal{T}$ is a subset of edges $c_t \subset
E(t)$. It becomes an admissible cut if and only if along a path
from the root to any of the leaves of the tree $t$, one encounters
at most one element of $c_t$. By removing the subset $c_t$ of
edges, each admissible cut $c_t$ produces a forrest of pruned
trees, denoted by $P_{c_t}$. The remaining part, which is a single
rooted tree linked to the original root, is denoted by $R_{c_t}$.
We exclude the cases, where $c_t=\emptyset$, such that
$R_{c_t}=t,\; P_{c_t}=\emptyset$ and the full cut, such that
$R_{c_t}=\emptyset,\; P_{c_t}=t$. The rooted tree algebra
$\mathcal{H}_{\mathcal{T}}$ is equipped with a bialgebra structure
by defining the coproduct $\Delta: \mathcal{H}_{\mathcal{T}} \to
\mathcal{H}_{\mathcal{T}} \otimes \mathcal{H}_{\mathcal{T}}$ in
terms of all admissible cuts $C_t$ of a rooted tree $t$:
\begin{equation}
    \Delta(t) = t \otimes 1_\mathcal{T} + 1_\mathcal{T} \otimes t
                       + \sum_{c_t \in C_t} P_{c_{t}} \otimes R_{c_t}.
    \label{coprod1}
\end{equation}
For example,
\allowdisplaybreaks{
\begin{eqnarray}
 \Delta(\ta1) &=&\;\: \ta1  \otimes 1_\mathcal{T} + 1_\mathcal{T} \otimes \ta1
    \notag \\
 \Delta\big(
 \!\!\begin{array}{c}
                 \\[-0.4cm]\tb2 \\
               \end{array}\!\! \big) &=&
                         \begin{array}{c}
                           \\[-0.5cm]\tb2 \\
                         \end{array} \!\! \otimes 1_\mathcal{T}
                            + 1_\mathcal{T} \otimes \!\!\!
                         \begin{array}{c}
                           \\[-0.5cm]\tb2 \\
                         \end{array}
                               + \ta1 \otimes \ta1 \notag\\
 \Delta\Big(\!\!\begin{array}{c}
                   \\[-0.5cm]\tc3 \\
                 \end{array}\!\!\Big)    &=&
                            \begin{array}{c}
                             \\[-0.5cm]\tc3 \\
                            \end{array}\!\! \otimes 1_\mathcal{T}
                         + 1_\mathcal{T} \otimes \!\!\!
                         \begin{array}{c}
                           \\[-0.5cm]\tc3 \\
                         \end{array}\!\!
                          + \ta1 \otimes\!\!\!
                         \begin{array}{c}
                           \\[-0.5cm]\tb2 \\
                         \end{array} +
                         \begin{array}{c}
                           \\[-0.5cm]\tb2 \\
                         \end{array}\!\! \otimes\! \ta1 \label{treeEx} \\
 \Delta\big(\:\td31\big) &=& \!\! \begin{array}{c}
                                 \\[-0.5cm]\td31 \\
                              \end{array}\!\! \otimes 1_\mathcal{T}
                               + 1_\mathcal{T} \otimes\!\!\!
                              \begin{array}{c}
                                 \\[-0.5cm]\td31 \\
                              \end{array}\!\! +
                                  2\ta1 \otimes \!\!\!
                                       \begin{array}{c}
                                         \\[-0.5cm]\tb2 \\
                                       \end{array} + \ta1\ta1\otimes \ta1
                                       \notag \\
 \Delta\bigg(\!\! \begin{array}{c}
                   \\[-0.5cm]\te4 \\
                  \end{array}\!\!\bigg)
                     &=& \begin{array}{c}
                           \\[-0.5cm]\te4 \\
                         \end{array} \!\! \otimes 1_\mathcal{T}
                         + 1_\mathcal{T} \otimes \!\!\!
                         \begin{array}{c}
                           \\[-0.5cm]\te4 \\
                         \end{array}
                         + \ta1 \otimes \!\!\!
                         \begin{array}{c}
                           \\[-0.5cm]\tc3 \\
                         \end{array}
                            + \begin{array}{c}
                        \\[-0.5cm]\tb2 \\
                         \end{array} \!\! \otimes \!\!\!
                         \begin{array}{c}
                        \\[-0.5cm]\tb2 \\
                         \end{array} +
                         \begin{array}{c}
                        \\[-0.5cm]\tc3 \\
                         \end{array}\!\! \otimes \! \ta1 \notag
\end{eqnarray}}
Latter we will use the shortened notation $\Delta(t)=\sum_{(t)}
t_{(1)}\otimes t_{(2)}$. For products of trees we demand the
compatibility $\Delta(t_1 \dots t_n):=\Delta(t_1)\dots
\Delta(t_n)$.

{\begin{rmk}{\label{linMap1} It is important to notice that the
right hand side of $\Delta(t) \in \mathcal{H}_{\mathcal{T}}
\otimes \mathcal{H}_{\mathcal{T}}$ is linear. Therefore we can
write, for the $\mathbb{K}$-vector space $\mathbb{K}\,
\mathcal{T}$ spanned by $\mathcal{T}$,
\begin{equation}
  \label{linMap2}
  \mathbb{K}\,\mathcal{T} \xrightarrow{\Delta} \mathcal{H}_{\mathcal{T}}
  \otimes \mathbb{K}\, \mathcal{T}
\end{equation}
}\end{rmk}}

The counit $\epsilon: \mathcal{H}_{\mathcal{T}} \to \mathbb{K}$
simply maps the empty tree $1_{\mathcal{T}}$ to $1_{\mathbb{K}}
\in \mathbb{K}$ and the rest to zero
\begin{equation}
  \label{counit}
  \epsilon(t_1 \cdots t_n):= \begin{cases}
                             1_{\mathbb{K}} & t_1\cdots t_n = 1_\mathcal{T},\\
                             0 & \;t_1\cdots t_n \ne 1_{\mathcal{T}}.
                             \end{cases}
\end{equation}

\begin{thm} {\rm \cite{CKI,Kreimer1}}
The algebra $\mathcal{H}_{\mathcal{T}}$ equipped with the above
defined compatible coproduct $\Delta:\mathcal{H}_{\mathcal{T}} \to
\mathcal{H}_{\mathcal{T}} \otimes \mathcal{H}_{\mathcal{T}}$ and
counit $\epsilon: \mathcal{H}_{\mathcal{T}} \to \mathbb{K}$ forms
a connected, graded, commutative, non-cocommutative bialgebra. In
fact it is a Hopf algebra, with antipode
$S:\mathcal{H}_{\mathcal{T}}\to\mathcal{H}_{\mathcal{T}}$ defined
recursively by $S(1_\mathcal{T})=1_{\mathcal{T}}$,
$S(t)=-t-\sum_{c_t \in C_t} S(P_{c_{t}}) R_{c_t}$.
\end{thm}
We denote the restricted dual of $\mathcal{H}_{\mathcal{T}}$ by
$\mathcal{H}^*_{\mathcal{T}}$. It contains linear maps from
$\mathcal{H}_{\mathcal{T}}$ into $\mathbb{K}$. Let $A$ be a
commutative $\mathbb{K}$-algebra, we also consider the linear
functionals $\Hom(\mathcal{H}_{\mathcal{T}},A)$. Equipped with the
convolution product:
\begin{equation}
  f \star g := m_A(f \otimes g)\Delta:
   \mathcal{H}_{\mathcal{T}} \xrightarrow{\Delta}
   \mathcal{H}_{\mathcal{T}} \otimes \mathcal{H}_{\mathcal{T}}
   \xrightarrow{f \otimes g} A \otimes A
   \xrightarrow{m_A} A
   \label{conv}
\end{equation}
$\Hom(\mathcal{H}_{\mathcal{T}},A)$ becomes a non-commutative
$\mathbb{K}$-algebra with unit given by the counit $\epsilon$ of
$\mathcal{H}_{\mathcal{T}}$.

A linear map $\phi: \mathcal{H}_{\mathcal{T}} \to A$ is called a
{\bf character} if $\phi$ is an algebra homomorphism. We denote
the set of characters by $G_A:=char_{A}\mathcal{H}_{\mathcal{T}}$.
\begin{prop}
 \label{groupchar} The set of characters $G_A$ forms a group with
 respect to the convolution product (\ref{conv}). The inverse of
 $\phi\in char_{A}\mathcal{H}_{\mathcal{T}}$ is given by
 $\phi^{-1}:=\phi \circ S$.
\end{prop}

A linear map $Z: \mathcal{H}_{\mathcal{T}} \to A$ is called a {\bf derivation},
or {\bf infinitesimal character} if
  \begin{equation}
    Z(t_1t_2)=Z(t_1)\epsilon(t_2)+\epsilon(t_1)Z(t_2),
    \;\;t_i \in \mathcal{H}_{\mathcal{T}},\ i=1,2.
    \label{deriv}
  \end{equation}
  We have $Z(1_{\mathcal{T}})=0$. The set of infinitesimal characters is denoted
  by $g_A:=\partial char_{A}\mathcal{H}_{\mathcal{T}}$.
For any $Z \in \partial char_{A}\mathcal{H}_{\mathcal{T}}$ and $t
\in \mathcal{H}_{\mathcal{T}}$ of degree $\#(t)=n<\infty$, we have
$Z^{\star m}(t)=0$ for $m>n$.  This implies that the exponential
$\exp^{\star}(Z)(t):=\sum_{k \ge 0}\frac{Z^{\star k}}{k!}(t)$, $Z
\in \partial char_{A}\mathcal{H}_{\mathcal{T}}$, is finite, ending
at $k=\#(t)$.
Given the explicit base of rooted trees generating
$\mathcal{H}_{\mathcal{T}}$, the $A$-module of derivations $\partial
char_{A}\mathcal{H}_{\mathcal{T}}$ is generated by the
dually defined infinitesimal characters, indexed by rooted trees
\begin{equation}
  Z_{t}(t') := \delta_{t,t'}1_{\mathbb{K}},\ t,t'\in \mathcal{T}.
  \label{Z}
\end{equation}

\begin{prop}
The set $g_A=\partial char_{A}\mathcal{H}_{\mathcal{T}}$
defines a Lie algebra when equipped with the
commutator:
\allowdisplaybreaks{
\begin{eqnarray}
    [Z_{t'},Z_{t''}] &:=& Z_{t'} \star Z_{t''} - Z_{t''} \star Z_{t'}     \nonumber\\
                     &=&  \sum_{t \in \mathcal{T}}
                                  \big( n(t',t'';t)-n(t'',t';t) \big)Z_t,
   \label{Liebra}
\end{eqnarray}}
where the $n(t',t'';t) \in \mathbb{N}$ denote so-called section
coefficients, which count the number of single admissible cuts,
$|c_t|=1$, such that $P_{c_{t}}=t'$ and $R_{c_t}=t''$.
\end{prop}

Ultraviolet divergencies demand for a regularization of the theory by
replacing the base field $\mathbb{K}$ in
$\mathcal{H}^*_{\mathcal{T}}$ as target space, by a commutative,
unital Rota--Baxter $\mathbb{K}$-algebra $(A,R)$. As
an example we mentioned earlier the commutative Rota--Baxter
algebra of Laurent series
$\mathcal{A}:=\mathbb{C}[\varepsilon^{-1},\varepsilon]]$.

The following theorem recalls the Birkhoff decomposition of Hopf
algebra characters from the work of Connes and Kreimer in the
context of complete filtered Rota--Baxter algebras.
\begin{thm} {\rm \cite{EGK1}} \label{SpitzerpQFT}
Let $\mathcal{H}$ be the connected graded Hopf algebra of rooted
trees, $\mathcal{H}_{\mathcal{T}}$ or Feynman graphs,
$\mathcal{H}_{\mathcal{F}}$, and let $(A,R)$ be a commutative
Rota--Baxter algebra with Rota--Baxter operator $R$ of weight $1$.
Let $\mathcal{A}:=Hom(\mathcal{H},A)$. Define a decreasing
filtration $\{\mathcal{A}_n\}_n$ on $\mathcal{A}$ in duality to
the increasing filtration on $\mathcal{H}$ by grading. Define
$\mathcal{R}:\mathcal{A}\to \mathcal{A}$ by $\mathcal{R}(f)=R\circ
f$, $f\in \mathcal{A}$. For a regularized character $\phi\in G_A$,
let
\begin{equation}
 \phi=\phi_-^{-1}\star \phi_+
 \label{CKBirkhoff}
 \end{equation}
be the algebraic Birkhoff decomposition of
Connes--Kreimer~\cite{CKII} that gives the counter term $\phi_-$
and renormalization $\phi_+$.
\begin{enumerate}
\item The triple $(\mathcal{A},\mathcal{R},\{\mathcal{A}_k\}_k)$
is a complete Rota--Baxter algebra under convolution~(\ref{conv}).
\item The decomposition (\ref{CKBirkhoff}) is the factorization of
$\phi$ in Theorem~\ref{thm:Atkinson}, with $\phi_{-} \in \epsilon
+ \mathcal{R}(\mathcal{A}_1)$ and $\phi_{+}^{-1} \in \epsilon +
\tilde{\mathcal{R}}(\mathcal{A}_1)$. For idempotent Rota--Baxter
map $R$ we have uniqueness of the factorization.
\item Explicitly we have \allowdisplaybreaks{
\begin{eqnarray*}
    \phi_{-}
    =\exp^\star\big(-\mathcal{R}\big(\chi(\log^{\star}(\phi))\big)\big),\qquad
    \phi^{-1}_{{+}} = \exp^\star\big(-\tilde{\mathcal{R}}\big(\chi(\log^{\star}(\phi))\big)\big)
\end{eqnarray*}}
where $Z_\alpha:=\log^\star (\phi)\in g_A$ with $\log^\star$ and
$\exp^\star$ defined by convolution product~(\ref{conv}), and
$\chi$ is the BCH-recursion (\ref{BCHrecursion}). \item Further,
$\phi_{-}$ and $\phi^{-1}_{+}$ solve Bogoliubov's recursions, i.e.
item (1) respectively item (2) of Theorem \ref{complRBo} on
Spitzer's identity, for $\phi-\epsilon \in \mathcal{A}_1$:
\allowdisplaybreaks{
\begin{eqnarray*}
  \begin{array}{cccccc}
 \phi_{-} \!\!&=&\hspace{-2.5cm}\epsilon - \mathcal{R}\big(\phi_{-}\star(\phi-\epsilon)\big)

   &\;\;\;\;\;\phi^{-1}_{+}\!\! &=&
       \hspace{-1.8cm} \epsilon - \tilde{\mathcal{R}}\big((\phi-\epsilon) \star \phi^{-1}_{+}\big)
       \\[0.3cm]
   &=&\epsilon +
              \mathcal{R}\big(\exp^{\star_{\mathcal{R}}}\big(-\chi(\log^{\star}(\phi))\big)-\epsilon\big)
           &\;\;\; &=            & \epsilon -
         \tilde{\mathcal{R}}\big(\exp^{\star_{\mathcal{R}}}\big(\chi(\log^{\star}(\phi))\big)-\epsilon\big)
  \end{array}
 \label{Bogoliubov}
\end{eqnarray*}}
\end{enumerate}
\end{thm}
\begin{prop} \label{bogoChar} {\rm \cite{EGK1}}
                               Let $\phi \in G_A$ be a Feynman rules character,
                               with $Z_\alpha:=\log^\star (\phi)\in g_A$ . The map
\begin{equation}
  \label{doubleExp}
    b[\phi] :=\exp^{{\star}_\mathcal{R}}\big(-\chi(Z_{\phi})\big)-\epsilon
\end{equation}
  defined in terms of the exponential with respect to the double Rota--Baxter product
  represents Bogoliubov's preparation map, also known under name $\bar{R}$-operation.
\end{prop}
Here $\exp^{{\star}_\mathcal{R}}$ denotes the exponential map with
respect to the Rota--Baxter double product (\ref{double}) defined
in the Rota--Baxter algebra
$(\mathcal{A},\mathcal{R},\{\mathcal{A}_k\}_k)$.


\subsection{Matrix representation of linear functionals}
\label{matrixRep}

Recall that a Hopf algebra $\mathcal{H}$ is called filtered if
there are $\mathbb{K}$-subvector spaces $\mathcal{H}^{(n)}$, $n
\geq 0$ of $\mathcal{H}$ such that
\begin{enumerate}
\item $\mathcal{H}^{(n)} \subseteq \mathcal{H}^{(n+1)}$;
\item $\bigcup_{n\geq 0} \mathcal{H}^{(n)} = \mathcal{H}$;
\item $\mathcal{H}^{(p)} \mathcal{H}^{(q)}\subseteq \mathcal{H}^{(p+q)}$;
\item $\Delta(\mathcal{H}^{(n)}) \subseteq \bigoplus_{p+q=n}
\mathcal{H}^{(p)}\otimes \mathcal{H}^{(q)}.$
\end{enumerate}
$\mathcal{H}$ is called connected, if in addition
$\mathcal{H}^{(0)}=\mathbb{K}$. Then for any $x \in
\mathcal{H}^{(n)}$, we have
\begin{equation}
   \tilde{\Delta}(x):=\Delta(x) - x\otimes 1 - 1\otimes x \in
       \oplus_{p+q=n,p>0,q>0} \mathcal{H}^{(p)}\otimes \mathcal{H}^{(q)}
   \label{eq:conn}
\end{equation}

\begin{defn}
A subset $X$ of $\mathcal{H}$ is called a {\bf{(left) subcoset}}
if $X$ is $\mathbb{K}$-linearly independent and if $\mathbb{K}\,
X$ is a left subcomodule of $\mathcal{H}$. A subcoset $X$ is
called {\bf{filtration ordered}} if $X$ is given an order that is
compatible with the order from the filtration of $\mathcal{H}$. In
other words, if $i \leq j$ and $x_i$ is in $\mathcal{H}_n$, then
$x_j$ is in $\mathcal{H}_n$. A subcoset $X$ is called a {\bf{
1-subcoset}} if $1_\mathcal{H}$ is in $X$.
\end{defn}

It is clear that $X$ is a left subcoset of $\mathcal{H}$ means
that the coproduct $\Delta$ on $\mathcal{H}$ restricts to
$\Delta_X: X \to \mathcal{H} \otimes X$.  In other words, for any
$x\in X$, with Sweedler's notation $\Delta_X(x)=\sum_{(x)}
x_{(1)}\otimes x_{(2)},$ we have $x_{(2)}\in \mathbb{K}\,X$.

This definition reflects the combinatorial character of certain
filtered Hopf algebras. Objects, mostly of graphical type, with
specific substructures are disentangled. For an example see the
coproduct example of the Feynman graph, \scalebox{0.5}{\coprod2},
in subsection \ref{FeynTrees}. The two subgraphs
\scalebox{0.5}{\l2cop} and \scalebox{0.5}{\r2cop} are replaced by
a vertex, \scalebox{1.1}{\vertex} , with the cograph
\scalebox{0.7}{\c2cop} as a result.
\medskip

Our matrix representation applies to any connected filtered Hopf
algebra $\mathcal{H}$ with a comodule. It naturally generalizes
the classical construction of modules from
comodules~\cite{Manchon,Sw}.

\begin{exams}{\rm
Examples of such Hopf algebras with a left 1-subcoset $X$ include
\begin{enumerate}
 \item the Hopf algebra $\mathcal{H}_\mathcal{T}$ of rooted trees
       with $X=\mathcal{T}$ being the set of rooted trees.
       See Remark~\ref{linMap1};
 \item the Hopf (sub)algebra of ladder trees $\mathcal{H}_{\ell d} \subset \mathcal{H}_\mathcal{T}$
       with $X=\ell d \subset \mathcal{T}$ being the set of ladder trees, i.e. trees whose
       vertices (except the root vertex) have only one incoming and at most one outgoing edge;
 \item the Hopf algebra $\mathcal{H}_{\mathcal{F}}$ of Feynman graphs with $X=\mathcal{F}$ being the set of
       one particle irreducible Feynman graphs. This follows from a remark
       similar to Remark~\ref{linMap1};
 \item one of the above Hopf algebras with $X$ being the first $n$ ($ n \geq 1$) elements
       in the subset there, with an ordering compatible with the filtration of the Hopf algebra.
\end{enumerate}}
\end{exams}


\subsubsection{The representation anti-homomorphism map} \label{coprodRepMap}

Let $A$ be a commutative $\mathbb{K}$-algebra. Denote
$A\,X=A\otimes \mathbb{K} X$ which is a free $A$-module with basis
$X$. For a Hopf algebra $\mathcal{H}$ with a left 1-subcoset $X$,
we will define a linear map $\Psi_{X,A}: \Hom(\mathcal{H},A) \to
\End(A X)$, eventually giving rise to a natural representation of
$\Hom(\mathcal{H},A)$ in terms of upper triangular matrices with
entries in $A$. In the following the subscript $X$ will be
suppressed when possible.
\begin{defn} \label{coprodRep}
Let $\mathcal{H}$ be a Hopf algebra with a filtration ordered left
1-subcoset $X$. $A$ is a commutative $\mathbb{K}$-algebra. The
linear map $\Psi_{A,X}: \Hom(\mathcal{H}, A) \to \End(A\,X)$ is
defined by taking the composition
\begin{equation}
  \Psi_{A,X}[f]: A\otimes \mathbb{K}X
  \xrightarrow{\id_A\otimes \Delta} A\otimes \mathcal{H} \otimes \mathbb{K}X
   \xrightarrow{\id_A\otimes f \otimes \id_{\mathbb{K}X}} A\otimes A \otimes \mathbb{K}X
   \xrightarrow{m_A\otimes \id_{\mathbb{K}X}} A\otimes \mathbb{K} X
  \label{eq:psi1}
\end{equation}
for $f\in \Hom(\mathcal{H},A)$.
\end{defn}
In particular, for $x\in \mathbb{K}X$ with $\Delta(x)=\sum_{(x)}
x_{(1)}\otimes x_{(2)}$, we have
\begin{equation}
   \Psi_A[f](x)=\sum_{(x)} f(x_{(1)})x_{(2)}.
    \label{eq:short}
\end{equation}
This justifies the short-hand notation $f\star
\id_{\mathbb{K}X}$ for the composition~(\ref{eq:psi1})
defining $\Psi_A[f]$.

As an example, consider the Hopf algebra $\mathcal{H}_\mathcal{T}$
of rooted trees with $X=\mathcal{T}$. Using the notation in
(\ref{coprod1}), we have, for $f\in
\Hom(\mathcal{H}_\mathcal{T},A)$ and $t\in \mathcal{T}$,
\begin{equation}
  \Psi_A[f](t) = f(1_\mathcal{T})t + f(t)1_\mathcal{T} + \sum_{c_t \in C_t} f(P_{c_t})R_{c_t}
  \label{linRep}
\end{equation}
which is in $A\,\mathcal{T}$.


\subsubsection{Coproduct matrix for filtered Hopf algebras}

With the natural basis $X$ of $A\,X$, the map $\Psi_A$ gives a
matrix representation of $\Hom(\mathcal{H},A)$. Explicitly, fix a
linear order $\{x_i\}_{i\geq 1}$ of the left 1-subcoset $X$ that
is compatible with the filtration of the Hopf algebra
$\mathcal{H}$. Then the coproduct $\Delta$ of $\mathcal{H}$
restricted to $X$ writes
\begin{equation}
 \Delta(x_j)=\sum_{i=1}^\infty X_{ij} \otimes x_i
 \label{eq:coprod}
 \end{equation}
for uniquely determined $X_{ij}$ in $\mathcal{H}$. Note the order
of $i$ and $j$. For all $i<j$, $\#(X_{ij}) < \#(x_j)$. This leads
to the
\begin{defn} \label{coprodMatrix}
             Let $\mathcal{H}$ be a filtered Hopf algebra.
             Fix a linear order $\{x_i\}_{i\geq 1}$
             of the filtration ordered left 1-subcoset $X$.
             We define the $|X| \times |X|$ matrix
$$
   M_\mathcal{H}:=M_{\mathcal{H},X}:=(X_{ij})
$$
in $\mathcal{M}_{|X|}(\mathcal{H})$, called the {\bf{coproduct
matrix of $\mathcal{H}$}} (with respect to $X$).
\end{defn}
Then $M_\mathcal{H}$ is upper triangular since $\mathcal{H}$ is
filtered. For a different choice of basis, we get conjugate
matrices. Under a fixed ordering $(x_1,\cdots,x_n,\cdots)$ of $X$,
we obtain an isomorphism
$$
  A\, X\to A^{|X|}
$$
sending $x_n$ to the unit column vector $\left |x_n\right >$ with
1 in the $n$-th entry and zero elsewhere. Here we use the familiar
bra-ket notation to denote abstract vectors as column vectors. We
likewise obtain the isomorphism
$$
  \End(AX) \to \mathcal{M}^u_{|X|}(A)
$$
sending $\Psi_A[f]$ to the upper triangular matrix
\begin{equation}
 \label{matphi}
 \widehat{f}=(f_{ij}):= \big(f(X_{ij})\big).
\end{equation}
We further have an isomorphism
$$
  \Hom(A\,X, A) \to A^{|X|}
$$
sending the dual basis $x_n^*$, defined by
$x_n^*(x_m)=\delta_{n,m}$, to the row vector $\left<x_n\right |$
with 1 in the $n$-th entry and zero elsewhere. The image of  $f\in
\Hom(A\,X,A)$ under this isomorphism is denoted by $\left<f\right
|$. So we have
$$
  \left <f\right |=(f(x_1),\cdots,f(x_n),\cdots).
$$
We will often use these isomorphisms as identifications when there
is no danger of confusion. In particular, $\widehat{f}$ is often
identified with $\Psi[f]$.

As an illustrating example we consider again the Hopf algebra
$\mathcal{H}_\mathcal{T}$ to rooted trees with a truncated $X$
being
\begin{equation}
\mathcal{T}_{(6)}:=
 \left\{e_1:= 1_\mathcal{T},\
e_2:= \ta1,\
e_3:=\!\! \begin{array}{c} \tb2 \\ \end{array}\!\!,\
e_4:=\!\!\begin{array}{c} \tc3 \\ \end{array}\!\!,\
e_5:=\!\!\!\begin{array}{c} \td31\ \\ \end{array}\!\!\!,\
e_6:= \!\!\begin{array}{c} \te4 \\ \end{array}\!\!
\right\} \label{6space}
\end{equation}
with the displayed linear order.
Then we have the corresponding column vectors
\begin{equation}
\left\{ \left|1_\mathcal{T}\right>,\
\left|\ta1\right>,\
 \big|\!\!\!\begin{array}{c} \tb2 \\ \end{array}\!\!\big>,\
\left|\!\!\begin{array}{c} \tc3 \\ \end{array}\!\!\right>,\
\big|\!\!\!\begin{array}{c} \td31 \\ \end{array}\!\!\!\big>,\
 \left|\!\!\begin{array}{c} \te4 \\ \end{array}\!\!\right>
\right\}
\end{equation}
In particular,
$$
  \left |1_\mathcal{T}\right >=\left (\begin{array}{c} 1 \\0\\ \vdots\\ 0\end{array} \right )
$$
and
$$
  \left < 1_\mathcal{T} \right | = (1,0,\cdots,0).
$$

{}From the coproducts in (\ref{treeEx}), we obtain the truncated
unipotent coproduct matrix from Definition \ref{coprodMatrix}
\begin{equation}
 \label{coprodMat6}
 M_\mathcal{H} = \left(
  \begin{array}{cccccc}
   1_\mathcal{T}& e_2           & e_3          & e_4           & e_5           & e_6 \\
   0               & 1_\mathcal{T} & e_2          & e_3           & e_2e_2        & e_4 \\
   0               & 0                & 1_\mathcal{T}& e_2           & 2 e_2         & e_3 \\
   0               & 0                & 0               & 1_\mathcal{T} & 0                & e_2 \\
   0               & 0                & 0               & 0                & 1_\mathcal{T} & 0 \\
   0               & 0                & 0               & 0                & 0                & 1_\mathcal{T}
  \end{array}
           \right)
\end{equation}
For an $f \in \Hom(\mathcal{H}_\mathcal{T},A)$ we then obtain its
representation $\widehat{f}_{(6)}$ in terms of an upper triangular
matrix by applying $f$ to $M_\mathcal{H}$ entry by entry. In each
column we just see the image of the left hand side of
$\Delta(e_i)$ under $f$.

For an infinitesimal character $Z \in g$ we find particularly
simple nilpotent matrices
\begin{equation}
  \label{a}
  \Psi[Z](t) =  Z(t)1_\mathcal{T} + \sum_{\substack{c_t \in C_t\\ |c_t|=1}} Z(t')t''.
\end{equation}
Here, due to relation (\ref{deriv}) we only need to consider
single admissible cuts, $|c_t|=1$. The sum on the right hand side
of (\ref{a}) therefore goes over all decompositions of the tree
$t$, resulting from the removal of exactly one edge. The tree $t'$
denotes the pruned subtree of tree $t$, and $t''$ the denotes the
tree, which is still connected to the root of $t$. For the
generators $Z_t$ we find
$$
  \Psi[Z_{t_1}](t_2)=\delta_{t_1,t_2}1_\mathcal{T} +
    \sum_{\substack{c_{t_2} \in C_{t_2}\\ |c_{t_2}|=1}} \delta_{t_1,t'_{2}}t''_{2}
    = \delta_{t_1,t_2}1_\mathcal{T} +
    \sum_{t} n(t_1,t;t_2)t.
$$
As an example, let us calculate
\begin{eqnarray*}
  \Psi[Z_{\ta1}]\big(\td31\big) &=& 2Z_{\ta1}(\ta1)\tb2=2\tb2\ , \qquad
  \Psi[Z_{\ta1}]\big(e_4\big) = Z_{\ta1}(\ta1)\tb2=\tb2\ , \qquad
  \Psi[Z_{\ta1}]\big(e_6\big) = Z_{\ta1}(\ta1)\tc3=\tc3 \\
  \Psi[Z_{\tc3}]\big(e_6\big) &=& Z_{\tc3}(e_4)\ta1=\ta1\ , \qquad
  \Psi[Z_{\tb2}]\big(e_4\big)  = Z_{\tb2}(\tb2)\ta1=\ta1\ , \qquad
  \Psi[Z_{\tb2}]\big(e_6\big)  = Z_{\tb2}(\tb2)\tb2=\tb2\ .
\end{eqnarray*}
 For the generators $Z_t$ of the Lie
algebra of derivations we find in this particular example the
following matrices. \allowdisplaybreaks{
\begin{align}
 \widehat{Z}_{\te4}= \left(
        \begin{array}{cccccc}
         0 & 0 & 0 & 0 & 0 & 1\\
         0 & 0 & 0 & 0 & 0 & 0\\
         0 & 0 & 0 & 0 & 0 & 0\\
         0 & 0 & 0 & 0 & 0 & 0\\
         0 & 0 & 0 & 0 & 0 & 0\\
         0 & 0 & 0 & 0 & 0 & 0
        \end{array}
  \right),\;
 \widehat{Z}_{\td31}= \left(
        \begin{array}{cccccc}
         0 & 0 & 0 & 0 & 1 & 0\\
         0 & 0 & 0 & 0 & 0 & 0\\
         0 & 0 & 0 & 0 & 0 & 0\\
         0 & 0 & 0 & 0 & 0 & 0\\
         0 & 0 & 0 & 0 & 0 & 0\\
         0 & 0 & 0 & 0 & 0 & 0
        \end{array}
  \right),\;
 \widehat{Z}_{\tc3}= \left(
        \begin{array}{cccccc}
         0 & 0 & 0 & 1 & 0 & 0\\
         0 & 0 & 0 & 0 & 0 & 1\\
         0 & 0 & 0 & 0 & 0 & 0\\
         0 & 0 & 0 & 0 & 0 & 0\\
         0 & 0 & 0 & 0 & 0 & 0\\
         0 & 0 & 0 & 0 & 0 & 0
        \end{array}
  \right),\;
\end{align}
\begin{align}
 \widehat{Z}_{\tb2}= \left(
        \begin{array}{cccccc}
         0 & 0 & 1 & 0 & 0 & 0\\
         0 & 0 & 0 & 1 & 0 & 0\\
         0 & 0 & 0 & 0 & 0 & 1\\
         0 & 0 & 0 & 0 & 0 & 0\\
         0 & 0 & 0 & 0 & 0 & 0\\
         0 & 0 & 0 & 0 & 0 & 0
        \end{array}
  \right),\;
  \widehat{Z}_{\ta1}= \left(
        \begin{array}{cccccc}
         0 & 1 & 0 & 0 & 0 & 0\\
         0 & 0 & 1 & 0 & 0 & 0\\
         0 & 0 & 0 & 1 & 2 & 0\\
         0 & 0 & 0 & 0 & 0 & 1\\
         0 & 0 & 0 & 0 & 0 & 0\\
         0 & 0 & 0 & 0 & 0 & 0
        \end{array}
  \right).
\end{align}}

For a character $\phi \in \Hom(\mathcal{H}_{\mathcal{T}},A)$, i.e.
an algebra homomorphism from $\mathcal{H}_{\mathcal{T}}$ to $A$,
$\phi(1_\mathcal{T})=1_A=1_{\mathbb{K}}$, we find the upper
triangular matrix with unit diagonal $\widehat{\phi} \in
\mathfrak{M}_6(A) \subset \mathcal{M}^u_6(A)$
\begin{equation}
 \widehat{\phi} = \left(
        \begin{array}{cccccc}
   1_{\mathbb{K}} & \phi(e_2)     & \phi(e_3)     & \phi(e_4)     & \phi(e_5)     & \phi(e_6) \\
   0              & 1_{\mathbb{K}}& \phi(e_2)     & \phi(e_3)     & \phi(e_2)^2   & \phi(e_4) \\
   0              & 0             & 1_{\mathbb{K}}& \phi(e_2)     & 2 \phi(e_2)   & \phi(e_3) \\
   0              & 0             & 0             & 1_{\mathbb{K}}& 0             & \phi(e_2) \\
   0              & 0             & 0             & 0             & 1_{\mathbb{K}}& 0         \\
   0              & 0             & 0             & 0             & 0             & 1_{\mathbb{K}}
  \end{array}
           \right)
\label{eq:phi6}
\end{equation}
The counit, which is a character by definition (\ref{counit}), is
represented by the unit matrix $\widehat{\epsilon}={\bf{1}}$. The
structure of the representation matrices shows that the row vector
$$
\left<1_\mathcal{T}\right|\widehat{\phi}= (1,0,0,0,0,0)\, \widehat{\phi}
$$
is
$$
  \left<\phi\right|:=(1_\mathbb{K},\phi(e_2),\cdots,\phi(e_6)).
$$

Returning to the general case, where $\mathcal{H}$ contains a
(filtration ordered left) 1-subcoset $X$, and $(A,R)$ is a
Rota--Baxter algebra, we have the following key property of
$\Psi_{A,X}$ that establishes the connection between the
Rota--Baxter algebras $\Hom(\mathcal{H},A)$ and
$\mathcal{M}_n^u(A)$. Remember that we suppressed the subindex $X$
if there is no danger of confusion.

\begin{thm} Let $\mathcal{H}$ be a connected filtered Hopf algebra
with a filtration ordered (left) 1-subcoset $X \subset
\mathcal{H}$. Let $A$ be a Rota--Baxter algebra (of weight 1).
\begin{enumerate}
 \item \label{it:anti} The linear map
   $$
     \Psi_A:=\Psi_{A,X}: \Hom(\mathcal{H},A) \to \mathcal{M}^u_{|X|}(A)
   $$
is an anti-homomorphism of Rota--Baxter algebras that is
continuous with respect to the topologies defined by the
filtrations on the filtered Rota--Baxter algebras. More precisely,
for any $m\geq 1$, there is $N\geq 1$, such that for all $k\geq N$
and $f\in \Hom(\mathcal{H},A)$ with $f(\mathcal{H}^k)=0$, we have
$\Psi_A[f] \in \mathcal{M}^u_{|X|}(A)_m$.
\item \label{it:row} The first row of $\widehat{f}:=\Psi_{A,X}[f] \in
\mathcal{M}^u_{|X|}(A)$ is
$\left<f\right|:=\left<1_X\right|\widehat{f}=(f(x))_{x \in X}$.
\item \label{it:inj} If $X$ is a generating set of the algebra
$\mathcal{H}$, then the map $\Psi_{A,X}$ restricts to an injective
map from the multiplicative group $G_A={\rm char}_A\mathcal{H}$ of
algebra homomorphism $\mathcal{H} \to A$ to $\End(A\,X)$.
\end{enumerate}
\label{thm:hom}
\end{thm}
\begin{proof}
(\ref{it:anti}) We denote the composition in $\End(A{X})$ by
concatenation. Let $f,g \in \Hom(\mathcal{H},A)$, and $x \in X$.
We will use Sweedler's notation of $\Delta(x)=x_{(1)}\otimes
x_{(2)}$ and the short hand notation $\Psi_A[f]=f\star \id_{AX}$
in (\ref{eq:short}). By the definition of $\Psi_A$, the
coassociativity of the coproduct $\Delta$ and the definition of
the convolution product $\star$, we have
\allowdisplaybreaks{
\begin{eqnarray}
  \Psi_A[f] \Psi_A[g](x) &=& (f \star \id_{AX}) \big( (g \star \id_{AX}) (x)\big) \nonumber\\
                     &=& (f \star \id_{AX}( g(x_{(1)}) x_{(2)})   \nonumber\\
                     &=& g(x_{(1)}) \big( f(x_{(2)(1)}) x_{(2)(2)}\big)              \nonumber\\
                     &=& \big( g(x_{(1)(1)}) f(x_{(1)(2)})\big) x_{(2)}           \label{coas}\\
                     &=& (g \star f)(x_{(1)})\: x_{(2)}                  \notag  \\
                     &=& \big((g \star f)\star \id_{AX}\big)(x)   \notag      \\
                     &=& \Psi_A[g \star f](x) \notag
\end{eqnarray}}
This proves the anti-homomorphism property.

Further, $\mathcal{R}$ is the Rota--Baxter map on
$\Hom(\mathcal{H},A)$, and we let $\widehat{\mathcal{R}}$ denote
the Rota--Baxter operator on $\mathcal{M}^u_{|X|}(A)$ by acting
entry-wise. Let $x \in X$ with $\Delta(x)=x_{(1)}\otimes x_{(2)}$.
We have, for $f \in \Hom(\mathcal{H},A)$, \allowdisplaybreaks{
\begin{align*}
\Psi_{A,X}[\mathcal{R}(f)](x)&= \mathcal{R}(f)(x_{(1)}) x_{(2)}\\
    &= R\big(f(x_{(1)})\big ) x_{(2)}\\
    &= \widehat{\mathcal{R}}(f(x_{(1)})x_{(2)})\\
    &= \widehat{\mathcal{R}}\big(\Psi_{A,X}[f]\big)(x).
\end{align*}}
This proves that $\Psi_{A,X}$ is compatible with the Rota--Baxter
operators.

The continuity is verified using the fact that the linear order on
$X$ is compatible with the filtration of $\mathcal{H}$.

(\ref{it:row})\ Since $X$ is assumed to contain $1=1_X$ and any
linear order of $X$ is assumed to be compatible with the
filtration of $\mathcal{H}$, we have
$$ X=(x_1:=1,x_2,\cdots)$$
for any choice of such linear ordering. So by Equation~(\ref{eq:conn}), we
have
$$\Delta(x_j)=x_j\otimes 1 + \sum_{i\geq 2} X_{ij}\otimes x_i.$$
Therefore the first row vector of the coproduct matrix
$M_\mathcal{H}$ is $(x_1,x_2,\cdots)$. So the first row of
$\widehat{f}$ is $(f(x_1),f(x_2),\cdots)$.

(\ref{it:inj})\ Suppose $\Psi_A[f]=0$. Then $\widehat{f}=0$. By item
(\ref{it:row}), $f(x)=0$ for all $x \in X$. Since $f: \mathcal{H}
\to A$ is an algebra homomorphism and $X$ is an algebra generating
set of $\mathcal{H}$, we have $f=0$. This proves the injectivity.
\end{proof}


\subsection{Renormalization of Feynman rules matrices}

We now consider the Hopf algebra $\mathcal{H}_{\mathcal{F}}$ of
Feynman graphs. Let $\mathcal{F}$ be the set of (equivalence
classes of) Feynman graphs with a fixed linear ordering
$\Gamma_1:=1, \Gamma_2,\cdots$ compatible with the filtration
$\mathcal{H}_\mathcal{F}^{(n)}$ of $\mathcal{H}_{\mathcal{F}}$.
Then with $X$ being either $\mathcal{F}$ or the subset
$\mathcal{F}_n$ of the first $n$-elements in $\mathcal{F}$, the
results of last subsection apply.

With the representation of a regularized Feynman character
$\phi:\mathcal{H}_\mathcal{F} \to A$ by an upper triangular matrix
obtained in Theorem~\ref{thm:hom}, we can apply the complete
filtered Rota--Baxter algebra structure on the matrices to
decompose the representation $\widehat{\phi}$, via Corollary
\ref{matdecomp}, into the inverse of the counter term matrix
$\widehat{\phi}_{-}$ and the renormalized matrix
$\widehat{\phi}_{+}$, giving rise to an analog of
Connes--Kreimer's algebraic Birkhoff decomposition for regularized
Feynman characters. Then by Theorem~\ref{thm:hom}.\ref{it:inj},
these later matrices recover the counter term and renormalization
of the regularized Feynman character~$\phi$.

In light of the anti-homomorphism $\Psi_{A}$ in
Theorem~\ref{thm:hom}, we expect that $\Psi_{A}$ exchanges the
order of the Birkhoff decomposition of $\phi$ in
(\ref{CKBirkhoff}). The precise relation between these two
decompositions is provided by the following theorem by applying
Theorem~\ref{thm:hom} and Proposition~\ref{pp:relation}. We will
use $\mathcal{R}$ to denote both, the Rota--Baxter map of
$\Hom(\mathcal{H}_{\mathcal{F}},A)$ and $\mathcal{M}_n^u(A)$,
since from the context it will be clear which one is used.

\begin{thm}
Let $\phi$ be a regularized character from
$\mathcal{H}_{\mathcal{F}}$ to $(A,R)$ and $Z_{\phi}=\log^\star
(\phi) \in g_A$. For a fixed $1 \leq n \leq \infty$, let
$X=\mathcal{F}_n$ be the first $n$ Feynman graphs and let
$\widehat{\phi} \in \mathfrak{M}_n(A) \subset \mathcal{M}_n^u(A)$ be
the upper triangular matrix representation of $\phi$ given by
(\ref{matphi}). Let
$$
  \phi=\phi_-^{-1} \star \phi_+
$$
be the Birkhoff decomposition of $\phi \in G_A$ in
(\ref{CKBirkhoff}). Let $\widehat{(\phi_-)}$ and
$\widehat{(\phi_+)}$ be the matrix representations of $\phi_- \in
G_A$ respectively $\phi_+ \in G_A$.
\begin{enumerate}
\item
Then $\widehat{(\phi_-)}$ is the unique solution of the equation
\begin{equation}
   \label{counterterm}
   \beta = {\bf{1}} - {\mathcal{R}}\big((\widehat{\phi}-{\bf{1}})\: \beta \big)
\end{equation}
explicitly given by
$$
  \widehat{(\phi_-)} = \exp\left(-{{\mathcal{R}}}(\bar{\chi} (\widehat{Z}_\phi)\right),
$$
and $\widehat{(\phi_+)}^{-1}$ is the unique solution of the
equation
\begin{equation}
  \label{renterm}
  \beta' = {\bf{1}} - \tilde{{\mathcal{R}}}(\beta' (\widehat{\phi}-{\bf{1}}))
\end{equation} explicitly given by
$$
  \widehat{(\phi_+)}^{-1}= \exp\left(-\tilde{{\mathcal{R}}}(\bar{\chi} (\widehat{Z}_\phi)\right).
$$
Here $\bar{\chi}$ is the BCH-recursion defined in
equation~(\ref{BCHrecursiono}).
\item $\widehat{\phi}$ factorizes as
\begin{equation}
  \label{opBirkhoff}
  \widehat{\phi} =  \widehat{(\phi_+)}\widehat{(\phi_-)}^{-1}.
\end{equation}
\item \label{it:mainrow} The first row vector of
$\widehat{(\phi_\pm)}$ is $\big(\phi_\pm(\Gamma)\big)_{\Gamma \in
\mathcal{F}_n}$. This can be summarized by the linear
renormalization matrix-vector equation for
$\left<\phi_{+}\right|:=\left<1_\mathcal{F}\right|\widehat{(\phi_+)}$
and $\left<\phi\right|:=\left<1_\mathcal{F}\right|\widehat{\phi}$:
\begin{equation}
\label{MatVectEq}
                 \left<\phi_{+}\right|=\left<\phi\right|\:\widehat{(\phi_-)},
\end{equation}
following from the matrix Birkhoff decomposition in
(\ref{opBirkhoff}).
\end{enumerate}
\label{thm:main}
\end{thm}

\begin{rmk}\label{inversRen}{\rm{The renormalized matrix $\widehat{(\phi_+)}$ uniquely
solves the equation
\begin{equation}
 \label{renterm2}
  c = {\bf{1}} - \tilde{{\mathcal{R}}}\big((\widehat{\phi}^{{}\ -1}-{\bf{1}})\ c\big),
\end{equation}
where we used Propositions \ref{opposite} and \ref{inverse}.
Instead of inverting $\widehat{(\phi_+)}^{-1} \longrightarrow
\widehat{(\phi_+)}$ at the end of calculation, here we first
invert the Feynman rules character matrix $\widehat{\phi}$, which
we can do simply and efficiently using the inverse coproduct
matrix, see below.}}
\end{rmk}


\subsection{Summary/Algorithm and examples}

We now discuss applications of our matrix approach to
renormalization in Theorem~\ref{thm:main}. One advantage of the
matrix approach lies in the efficiency to calculate the matrices
$\widehat{(\phi_{-})}$ and $\widehat{(\phi_{+})}$ of the matrix
Birkhoff factorization directly in the matrix algebra, see
Corollary~\ref{matdecomp}, using the recursions of
Theorem~\ref{thm:matrix}.

We outline an algorithm which is followed by examples.


\subsubsection{Calculation of the coproduct matrix and its inverse}

Let $X:=\mathcal{T}$ be the set of rooted trees (or Feynman
graphs) and fix an ordering of $\mathcal{T}$ that is compatible
with the grading of $\mathcal{T}$,
$t_1:=1_{\mathcal{T}},t_2,\cdots, t_n,\cdots$. Then the coproduct
$\Delta$ on $\mathcal{H}_\mathcal{T}$ is given~by
$$
  \Delta(t)=\sum_{(t)} t' \otimes t'',
$$
or in our ordering,
\begin{equation}
 \Delta(t_j)=\sum_{i=1}^j T_{ij} \otimes t_i
 \label{eq:coprod}
 \end{equation}
where $T_{ij}$ is in $\mathcal{H}_\mathcal{T}$, $t_i$ is in
$\mathcal{T}$, see Remark \ref{linMap1}, and
$\#(T_{ij})\leq\#(t_j)$. Define
$$
  M_\mathcal{H}=(T_{ij})
$$
to be the coproduct matrix of $\mathcal{H}_\mathcal{T}$.
$M_\mathcal{H}$ is an $\infty \times \infty$ upper triangular
matrix with entries in $\mathcal{H}_\mathcal{T}$ and unit on the
diagonal. The entries $T_{ij}$ can be obtained by the $B^+$
operator of Connes and Kreimer providing a recursive way to
calculate the coproduct. One can also consider the truncated
coproduct matrix by restriction to the subspace of the first $n$
graphs $t_1,\cdots,t_n$ and obtain a finite upper triangular
matrix in $\mathcal{M}^u_{n}(\mathcal{H})$.\smallskip

The direct calculation of the renormalization matrix
$\widehat{\phi_{+}}$ in Theorem~\ref{thm:main} respectively Remark
\ref{inversRen} demands an inversion of character matrices. In
Proposition \ref{groupchar} we stated the fact that the inverse of
a Hopf algebra character $\phi \in G$ is given by composition with
the antipode anti-homomorphism, $\phi^{-1}= \phi \circ S$. The
representation matrix of $\phi$ loosely speaking follows from
(\ref{matphi}), $\widehat{\phi}:= \phi \circ M_\mathcal{H}$. As we
want to calculate the inverse coproduct matrix,
$M^{-1}_\mathcal{H}=(T^{-1}_{ij})$, such that
$\widehat{\phi}^{-1}:= \phi \circ M^{-1}_\mathcal{H}$, we may take
another look at the coproduct matrix $M_\mathcal{H}$.

For this purpose we go back to Definition \ref{coprodRep}, where
we define the commutative $\mathbb{K}$-algebra $A:=\mathcal{H}$.
For a filtered Hopf algebra $\mathcal{H}$ with a filtration
ordered (left) 1-subcoset $X$ this provides us with an upper
triangular matrix representation of
$\Hom(\mathcal{H},\mathcal{H})$. The coproduct matrix thereby
represents the identity homomorphism $\id_{\mathcal{H}}:
\mathcal{H} \to \mathcal{H}$
\begin{equation}
   \label{copordMatRepresent}
   \Psi_{\mathcal{H}}[\id_{\mathcal{H}}]=\widehat{\id_{\mathcal{H}}}=M_\mathcal{H}.
\end{equation}
We suppressed the 1-subcoset $X\subset \mathcal{H}$ in the
notation.

As an upper triangular matrix, the inverse of $M_\mathcal{H}$
follows immediately in a recursive manner from Equation
(\ref{antipodeInverse}). But we want to make a little detour,
using the aforementioned representation point of view in
(\ref{copordMatRepresent}). This will provide us with a
non-recursive simple formula for calculating $M^{-1}_\mathcal{H}$.

The antipode $S$ in a Hopf algebra $\mathcal{H}$ is an
anti-homomorphism, defined as the solution of the equation $S
\star \id_{\mathcal{H}}=\epsilon=\id_{\mathcal{H}} \star S$. From
a convolution product point of view the antipode is the inverse of
the identity map, $S=\id^{\star -1}_{\mathcal{H}}$. For connected
filtered Hopf algebras the identity can simply be written as
$\id_{\mathcal{H}}=\exp^{\star}(\log^{\star}(\id_{\mathcal{H}}))$,
using the bijectivity of $\exp^{\star}$ and $\log^{\star}$, and
therefore we have
$S=\exp^{\star}(-\log^{\star}(\id_{\mathcal{H}}))$. We already
mentioned the trivial fact that every algebra is a Rota--Baxter
algebra, with Rota--Baxter operator pair $\id$ and
$\tilde{\id}=0$. The space $\End(\mathcal{H})$ is equipped with
two products, composition and convolution, both forming an
associative algebra. Spitzer's identity for the non-commutative
algebra $(\End(\mathcal{H}),\star)$ with Rota--Baxter map
$\id_\mathcal{H}:\End(\mathcal{H})\to\End(\mathcal{H})$ then
implies for the antipode
$$
  S=\exp^{\star}\left(-\id_{\mathcal{H}}\big( \log^{\star}(\epsilon
                                + (\id_{\mathcal{H}}-\epsilon))\big)\right)
$$
to be a solution of the equation $b= \epsilon -
\id_{\mathcal{H}}\big(b \star (\id_{\mathcal{H}}-\epsilon))$.
Explicitly
\begin{equation}
  \label{antipode}
  S = \epsilon - (\id_{\mathcal{H}}-\epsilon) +
  (\id_{\mathcal{H}}-\epsilon)\star(\id_{\mathcal{H}}-\epsilon) -
   (\id_{\mathcal{H}}-\epsilon)\star(\id_{\mathcal{H}}-\epsilon)\star(\id_{\mathcal{H}}-\epsilon)
                                                                                            + \cdots
\end{equation}
The BCH-recursion $\chi$ (\ref{BCHrecursion}) obviously does not
enter in this particular case as
$\widetilde{\id_{\mathcal{H}}}=0$. See~\cite{FG2} for the
geometric series ansatz. Applying the matrix representation
anti-homomorphism $\Psi_{\mathcal{H}}$, we find the {\bf{inverse
coproduct matrix}}
\begin{eqnarray}
  \label{antipodeMatrix}
  \Psi_{\mathcal{H}}[S] &=& \widehat{\id^{\star -1}_{\mathcal{H}}}=M^{-1}_\mathcal{H}\\
                        &=& {\bf{1}} - (M_\mathcal{H}-{\bf{1}}) +
                                             (M_\mathcal{H}-{\bf{1}}) (M_\mathcal{H}-{\bf{1}})  -
                            (M_\mathcal{H}-{\bf{1}}) (M_\mathcal{H}-{\bf{1}}) (M_\mathcal{H}-{\bf{1}})  +\cdots\\
                        &=& {\bf{1}}-\sum_{n>0}(M_\mathcal{H}-{\bf{1}})^{n}.
\end{eqnarray}
The inverse character matrix of $\phi \in G$ therefore is given by
\begin{equation}
  \label{inverseChar}
  \widehat{\phi}^{-1}=\phi \circ M^{-1}_\mathcal{H} = {\bf{1}}-
                                        \sum_{m>0}(\widehat{\phi}-{\bf{1}})^{m}.
\end{equation}
For a truncated coproduct matrix $M_\mathcal{H} \in
\mathcal{M}^u_n(\mathcal{H})$, $n<\infty$, the series on the right
breaks up at order $n$, $(\widehat{\phi}-{\bf{1}})^n=0$, as
$\widehat{\phi}-{\bf{1}} \in \mathcal{M}^u_n(A)_1$ is nilpotent.
In components, the formula for $\widehat{\phi}^{-1}$ is
\begin{equation}
 \label{inverseComponents}
  (\widehat{\phi}^{-1})_{ij}= -\widehat{\phi}_{ij}
              + \sum_{k=1}^{j-i-1}\:\sum_{i<l_1<l_2<\cdots<l_k<j}
                         (-1)^{k+1}
                         \widehat{\phi}_{il_1}\widehat{\phi}_{l_1l_2}
                         \dots \widehat{\phi}_{l_kl_j}.
\end{equation}


\subsubsection{Matrix renormalization by factorization}

Now let $\phi:\mathcal{H}_\mathcal{T} \to A$ denote a regularized
Feynman rules character with image in a Rota--Baxter algebra
$(A,R)$. Applying $\phi$ to $M_\mathcal{H}$ gives the {\bf{Feynman
rules matrix}}
$$
  \widehat{\phi}:= \phi( M_\mathcal{H})=\big (\phi(T_{ij})\big ).
$$

Let $\beta$ be the unique solution of the recursion
$$
  \beta={\bf 1}- {\mathcal{R}}\big((\widehat{\phi}-{\bf 1})\beta\big),
$$
as in Theorem~\ref{thm:main}. The matrix $\beta$ can be
effectively computed by Theorem~\ref{thm:matrix}.1. The first row
vector of $\beta$ is the counter term vector
$(\phi_-(t_1),\cdots,\phi_-(t_n),\cdots)$. Then the first row
vector of the matrix product $\widehat{\phi}\:\beta$ is the
renormalization vector, i.e., we have the linear renormalization
matrix-vector equation (\ref{MatVectEq}) of item (3) in
Theorem~(\ref{thm:main})
$$
\left<1_\mathcal{T}\right|\widehat{\phi}\:\beta
=\left<\phi\right|\:\beta=(\phi_+(t_1),\cdots,\phi_+(t_n),\cdots).
$$

Alternatively, let $\beta'$ be the unique solution of the recursion
$$
  \beta'={\bf 1}- \tilde{{\mathcal{R}}}(\beta'(\widehat{\phi}-{\bf 1})),
$$
as in Theorem~\ref{thm:main}, again effectively computable by
Theorem~\ref{thm:matrix}.2. Then find $\beta'{}^{-1}$ which can be
computed by Equation~(\ref{inverseChar}) (or
recursively~(\ref{antipodeInverse})). Then the first row of
$\beta'{}^{-1}$ is the renormalization vector
$$
 \left<1_\mathcal{T}\right| \beta'{}^{-1}=\left<\phi_{+}\right|
  =(\phi_+(t_1),\cdots,\phi_+(t_n),\cdots).
$$
Equivalently, using the inverse coproduct matrix to calculate the
inverse Feynman rules character
$$
  \widehat{\phi}{}^{-1}:= \phi( M^{-1}_\mathcal{H})=\big (\phi(T^{-1}_{ij})\big),
$$
we find directly the renormalized character matrix as solution of
the equation $c={\bf 1}
-\tilde{{\mathcal{R}}}\big((\widehat{\phi}{}^{-1}-{\bf 1}\big)\
c\big)$ of Remark \ref{inversRen}.


\subsubsection{Examples in $\varphi_{4dim}^{4}$-theory}
\label{FeymanExample}

This subsection serves to show how the above Hopf algebra
consideration nicely applies to standard Feynman graph
calculations of perturbative renormalization. In \cite{CKII}
Connes and Kreimer showed in full generality that the set of
Feynman graphs $\mathcal{F}$ for any perturbatively treated QFT
can be made into a combinatorial Hopf algebra
$\mathcal{H}_\mathcal{F}$ of the above type.

We will use a simplified version of $\varphi^4$-theory in four
dimensions as our Feynman graph toy model physics theory. A more
detailed and refined treatment can be found in the companion
paper~\cite{Echo}.

The Feynman graph Hopf algebra is denoted by
$\mathcal{H}_{\mathcal{F}}$. As regularization scheme we choose
dimensional regularization. So that the space of linear
functionals $\Hom(\mathcal{H}_{\mathcal{F}},A)$ contains maps into
the commutative Rota--Baxter algebra
$A:=\mathbb{C}[\varepsilon^{-1},\varepsilon]]$ with Rota--Baxter
map $R_{ms}$.

We work up to 3-loop order, by taking $X$ to be the set of graphs
\begin{equation}
 \mathcal{F}_{(4)} :=  \left\{
 e_1:= \left|1_\mathcal{F}\right>,\
 e_2:= \left|\newfish\right>,\
 e_3:= \big|\winecup\big>,\
 e_4:= \Big|\kite\Big>  \right\} \label{4space}
\end{equation}
identified with the corresponding column vectors. The graph \kite\
is obtained by substituting a wine-cup \winecup\ into the
divergent 1-loop graph \newfish.

The Feynman graphs we consider here are of purely combinatorial
type in the sense that the external legs are not decorated by
external structure, such as external momenta, spin indices, etc.
This frees us from symmetry consideration, which otherwise would
demand a bigger representation space $\mathcal{F}$.

The Feynman diagram \newfish\ has a primitive divergence and would
correspond to the one vertex tree \scalebox{2}{\ta1} decorated by
this graph. the wine-cup diagram \winecup\ contains exactly one
nested subdivergence of type \newfish\, and corresponds to the
ladder graph of length 2, \scalebox{1.1}{\tb2}, where the root and
leaf both are decorated by the graph \newfish. The coproducts of
these two graphs therefore are given in analogous forms to the
first two expressions in (\ref{treeEx}). The graph \kite\ contains
three nested subdivergences and correspondence to the ladder graph
of length 3, with each vertex decorated by \newfish, its coproduct
is given by
$$
 \Delta\Big(\kite \Big) = \kite \otimes 1_{{\mathcal{F}}}
                           + 1_{{\mathcal{F}}}\otimes \kite +
                           \newfish \otimes \winecup +
                           \winecup \otimes \newfish.
$$
Let $\phi \in G_A$ denote the Feynman rules for four dimensional
$\varphi^4$-theory in dimensional regularization, together with
minimal subtraction scheme, i.e. Rota--Baxter map $R:=R_{ms}$. The
corresponding coproduct matrix $M_{\mathcal{F}_{(4)}}$,
respectively character matrix $\Psi_A[\phi]=\widehat{\phi}$ are
given by \allowdisplaybreaks{
\begin{eqnarray}
  \widehat{\phi}:=\phi\circ M_{\mathcal{F}_{(4)}}=
\phi \circ \left(
        \begin{array}{cccc}
         1 & \newfish  & \winecup  & \kite     \\
         0 & 1         & \newfish  & \winecup  \\
         0 & 0         & 1         & \newfish  \\
         0 & 0         & 0         & 1
        \end{array}
            \right):=
  \left(
        \begin{array}{cccc}
         1 & \phi\big(\newfish\big) & \phi\big(\winecup \big) & \phi\Big(\kite\Big)\\
         0 & 1         & \phi\big(\newfish\big)               & \phi\big(\winecup\big)\\
         0 & 0         & 1                                    & \phi\big(\newfish\big)\\
         0 & 0         & 0                                    & 1
        \end{array}
  \right).\nonumber
\end{eqnarray}}
This matrix has the special property that along any subdiagonal we
find the same entry. This situation always appears if (and only
if) we deal with strictly nested diagrams only. In the picture of
decorated rooted trees this corresponds to the Hopf subalgebra of
ladder trees.

Using the counter term recursion (\ref{counterterm}) of
Theorem~\ref{thm:main} for $\widehat{\phi}$,
$$
   \beta={\bf{1}} - \mathcal{R}\left((\widehat{\phi}-{\bf{1}})\:\beta\right)
$$
and applying formula (\ref{barRec1}) for its solution, we find
\allowdisplaybreaks{\begin{eqnarray}
  \beta:=\left(
        \begin{array}{cccc}
         1 & -R\big(\phi\big(\newfish\big)\big) &\beta_{13} & \beta_{14}\\
         0 & 1         & -R\big(\phi\big(\newfish\big)\big) & \beta_{24}\\
         0 & 0         & 1         & -R\big(\phi\big(\newfish\big)\big) \\
         0 & 0         & 0         & 1
        \end{array}
  \right),\nonumber
\end{eqnarray}}
where
$$
\beta_{13}= -R\Big(\phi\big(\winecup \big)\Big)
            +R\bigg(\phi\big(\newfish\big)R\Big(\phi\big(\newfish\big)\Big)\bigg),
$$
$$
  \beta_{24}=\beta_{13},
$$
and \allowdisplaybreaks{
\begin{eqnarray}
\beta_{14}&=&-R\bigg(\phi\big(\kite\big)\bigg)
                                    +R\bigg(\phi\big(\winecup\big)
                                    R\Big(\phi\big(\newfish\big)\Big)\bigg)
                                    +R\bigg(\phi\big(\newfish\big)
                                    R\Big(\phi\big(\winecup\big)\Big)\bigg)\\
                             & & \hspace{2cm}\; \qquad \qquad - R\bigg(\phi\big(\newfish\big)
                                    R\bigg(\phi\big(\newfish\big)
                                    R\Big(\phi\big(\newfish\big)\Big)\bigg)\bigg)\\
                             &=& \phi_{-}\bigg(\kite\bigg).
\end{eqnarray}}
So in the end we have the following counter term matrix
\allowdisplaybreaks{
\begin{eqnarray*}
   \widehat{\phi}_{-}\!\!&=&\!\!
  {\small{\left(
        \begin{array}{cccc}
         1 & -R\Big(\phi\big(\newfish\big)\Big)
                              & -R\Big(\phi\big(\winecup \big)
                                - \phi\big(\newfish\big)R\Big(\phi\big(\newfish\big)\Big)\bigg)
                              & \phi_{-}\bigg(\kite\bigg)\\
         0 & 1                & -R\Big(\phi\big(\newfish\big)\Big)  &
                                 -R\bigg(\phi\big(\winecup\big)- \phi\big(\newfish\big)
                                    R\Big(\phi\big(\newfish\big)\Big)\bigg)\\
         0 & 0                & 1                             & -R\Big(\phi\big(\newfish\big)\Big)\\
         0 & 0                & 0                             & 1
        \end{array}
  \right)}}
\end{eqnarray*}}
giving the counter terms for the graphs \newfish, \winecup\ and
\kite\ in its first row. The renormalized character matrix
$\widehat{\phi_{+}}$ follows from the matrix Birkhoff
factorization (\ref{opBirkhoff}) in Theorem~\ref{thm:main},
$\widehat{\phi}_{+}=\widehat{\phi}\:\widehat{\phi}_{-}$. According
to the matrix-vector equation (\ref{MatVectEq}) of
Theorem~\ref{thm:main} we find
$$
 \left<\phi_{+}\right|=\left<1_\mathcal{F}\right|\widehat{\phi}_{+}
                      = \left<\phi\right|\widehat{\phi}_{-},
$$
yielding the renormalized amplitudes for the graphs \newfish,
\winecup, and \kite\ in its first row, which we write in
transposed form \allowdisplaybreaks{
\begin{eqnarray*}
  \left< \phi_+ \right|^{\top}\!\! \!\!&=&\!\!\!\!
     \left(\!\!\!\!
        \begin{array}{c}
         1 \\
         \phi\big(\newfish\big) - R\Big(\phi\big(\newfish\big)\Big)\\
          \phi\big(\winecup \big) - \phi\big(\newfish\big)R\Big(\phi\big(\newfish\big)\Big)
         -R\Big(\phi\big(\winecup \big)\Big)
          + R\bigg(\phi\big(\newfish\big) R\Big(\phi\big(\newfish\big)\Big)\bigg)\\
      \makebox{{\small{$  \left\{\!\!\!\!
               \begin{array}{l}
                                   \phi\Big(\kite\Big)
                                    -\phi\big(\winecup\big)
                                    R\Big(\phi\big(\newfish\big)\Big)
                                    -\phi\big(\newfish\big)
                                    R\Big(\phi\big(\winecup\big)\Big)
                                    +\phi\big(\newfish\big)
                                    R\Big(\phi\big(\newfish\big)
                                    R\Big(\phi\big(\newfish\big)\Big)\Big)\\
                                    -R\bigg(
                                    \phi\Big(\kite\Big)
                                    -\phi\big(\winecup\big)
                                    R\Big(\phi\big(\newfish\big)\Big)
                                    -\phi\big(\newfish\big)
                                    R\Big(\phi\big(\winecup\big)\Big)
                                    +\phi\big(\newfish\big)
                                    R\Big(\phi\big(\newfish\big)
                                    R\Big(\phi\big(\newfish\big)\Big)\Big)
                                    \bigg)\\
               \end{array}
        \!\!\!\right\} $}}}
        \end{array}
  \!\!\!\!\right)
\end{eqnarray*}}

Another example up to 3-loop order, including a graph with two
disjoint 1-loop subdivergences is provided by taking $X$ to be the
set of graphs
\begin{equation}
\mathcal{F}'_{(4)} :=  \left\{
e_1:= \left|1_{\mathcal{F}'}\right>,\
e_2:= \left|\newfish\right>,\
e_3:= \big|\winecup\big>,\
e_4:= \Big|\:\roll \Big> \! \right\} \label{4bspace}
\end{equation}
The new graph \roll\ is made of the two disjoint fish graphs
\newfish\ as subdivergences sitting inside of such a \newfish\
graph. Remember that our graphs carry no external structure. The
coproduct of this graph is given by
$$
 \Delta\Big(\roll \Big) = \roll \otimes 1_{\mathcal{F}'}
                           + 1_{\mathcal{F}'}\otimes \roll +
                           2\newfish \otimes \winecup +
                            \newfish \newfish \otimes \newfish.
$$
Let $\phi \in G_A$ be the Feynman rules character. The coproduct
matrix, respectively the character matrix are given by
\allowdisplaybreaks{
\begin{eqnarray}
  \widehat{\phi}:=\phi\circ M_{\mathcal{F}'_{(4)}}=
\phi \circ \left(
        \begin{array}{cccc}
         1 & \newfish  & \winecup  & \roll             \\
         0 & 1         & \newfish  & \newfish\newfish  \\
         0 & 0         & 1         & 2\newfish         \\
         0 & 0         & 0         & 1
        \end{array}
            \right):=
  \left(
        \begin{array}{cccc}
         1 & \phi\big(\newfish\big) &  \phi\big(\winecup\big)  & \phi\Big(\roll\Big) \\
         0 & 1                      &  \phi\big(\newfish\big)  & \phi\big(\newfish\big)^{2}\\
         0 & 0                      &  1                       & 2\phi\big(\newfish\big)\\
         0 & 0                      &  0                       & 1
        \end{array}
  \right).\nonumber
\end{eqnarray}}
In this example the only new counter term matrix entry we need to
calculate is position $(1,4)$ in $\widehat{\phi}_{-}$. Applying
formula (\ref{barRec1}) for its solution, we find
\allowdisplaybreaks{
\begin{eqnarray}
\beta_{14}&=&-R\bigg(\phi\Big(\roll\Big)\bigg)
                                    +R\bigg(\phi\big(\newfish\big)
                                     R\Big(\phi\big(\newfish\big)^2\Big)\bigg)
                                    +R\bigg(\phi\big(\winecup\big)
                                     R\Big(2\phi\big(\newfish\big)\Big)\bigg)\\
           & & \hspace{3cm} - R\bigg(\phi\big(\newfish\big)
                                    \:2R\Big(\phi\big(\newfish\big)
                                    R\Big(\phi\big(\newfish\big)\Big)\Big)
                                    \bigg).\nonumber
\end{eqnarray}}

The identity $R(a)^2=2R(aR(a))-R(a^2)$ following from (\ref{RB}),
and which is true only for commutative Rota--Baxter algebras,
immediately implies \allowdisplaybreaks{
\begin{eqnarray}
\beta_{14}&=&-R\bigg(\phi\Big(\roll\Big)\bigg)
                                    -R\bigg(\phi\big(\newfish\big)
                                     R\Big(\phi\big(\newfish\big)\Big)^2\bigg)
                                    +2R\bigg(\phi\big(\winecup\big)
                                     R\Big(\phi\big(\newfish\big)\Big)\bigg).
\end{eqnarray}}
Likewise for the renormalized expression we find in entry $(1,4)$
of the renormalized matrix $\widehat{\phi}_{+}$, respectively the
4th component of the vector $\left< \phi_+ \right|^{\top}$
\allowdisplaybreaks{
\begin{eqnarray}
  \label{BCexample}
  (\left< \phi_+ \right|^{\top})_4 &=& \phi\Big(\roll\Big)
                                    +\phi\big(\newfish\big)
                                     R\Big(\phi\big(\newfish\big)\Big)^2
                                    -2\phi\big(\winecup\big)
                                     R\Big(\phi\big(\newfish\big)\Big)\\
                                  & &  -R\bigg(\phi\Big(\roll\Big)
                                               +\phi\big(\newfish\big)
                                               R\Big(\phi\big(\newfish\big)\Big)^2
                                              -2\phi\big(\winecup\big)
                                               R\Big(\phi\big(\newfish\big)\Big)\bigg).\nonumber
\end{eqnarray}}


\subsection{More examples and comments on matrix factorization}

As another illustration, let us consider the case of the truncated
space $\mathcal{T}_{(6)}$ in (\ref{6space}) of undecorated rooted
trees. For a given regularized character $\phi:
\mathcal{H}_\mathcal{T}\to A$, we have the corresponding matrix
$\widehat{\phi}$ in Equation (\ref{eq:phi6}) which we record below
for easy reference.
\allowdisplaybreaks{
\begin{eqnarray}
\label{bigexample}
  \widehat{\phi}&=& \phi \circ M_\mathcal{H} \\
                &=& \left(
        \begin{array}{cccccc}
   1  & \phi(e_2)   & \phi(e_3) & \phi(e_4) & \phi(e_5)    & \phi(e_6) \\
   0  & 1               & \phi(e_2) & \phi(e_3) & \phi(e_2)^2  & \phi(e_4) \\
   0  & 0               & 1             & \phi(e_2) & 2 \phi(e_2)  & \phi(e_3) \\
   0  & 0               & 0             & 1             & 0                & \phi(e_2) \\
   0  & 0               & 0             & 0             & 1                & 0 \\
   0  & 0               & 0             & 0             & 0                & 1
  \end{array}
           \right),\nonumber
\end{eqnarray}}
where $M_\mathcal{H}$ is the coproduct matrix in
(\ref{coprodMat6}). Recall that the unit diagonal upper triangular
matrix
\begin{equation}
  \label{eins}
\beta:=\widehat{\phi_-}=
\exp\left(-\mathcal{R}(\bar{\chi}(\widehat{Z}_{\phi}))\right)={\bf{1}}+
        \mathcal{R}\left(\exp^{*_\mathcal{R}}\left(-\bar{\chi}(\widehat{Z}_{\phi})\right)
                               -{\bf{1}}\right)
\end{equation}
is solution to the equation (\ref{counterterm})
$$
   \beta ={\bf{1}} -\mathcal{R}\left((\widehat{\phi}-{\bf{1}})\:\beta\right).
$$
Remember that the second equality in (\ref{eins}) follows from
Proposition \ref{doubleExp} respectively its matrix
representation, with $\bar{\chi}$ in place of $\chi$, see Theorem
\ref{thm:main}. .

Equation  (\ref{counterterm}) is solved by the formula
(\ref{barRec1}), and we find
\allowdisplaybreaks{\begin{eqnarray*}
  \beta:=\left(
        \begin{array}{cccccc}
1& -R(\phi(e_2)) & \beta_{13} &\beta_{14} & \beta_{15} & \beta_{16} \\
0& 1    & -R(\phi(e_2)) &\beta_{24} &\beta_{25} &\beta_{26} \\
0&  0&   1 & -R(\phi(e_2))    & \beta_{35}  &\beta_{36}\\
0&  0  &0  &1  &0  &    \beta_{46}    \\
0&  0  &0  &0  &1  & \beta_{56} \\
0&  0  &0  &0  &0  &1
\end{array} \right )
\end{eqnarray*}}
where, in abbreviating $\phi(e_i)$ by $e_i$, $1\leq i\leq 6$,
\allowdisplaybreaks{
\begin{eqnarray*}
\beta_{13}&=& -R(e_3) + R\big(e_2R(e_2)\big),\\
\beta_{14}&=& -R(e_4) + R\big(e_2R(e_3)\big) +
                                        R\big(e_3R(e_2)\big)
                                               -R\big(e_2 R(e_2 R(e_2))\big),\\
\beta_{15}&=& -R(e_5) + R\big(e_2R(e_2e_2)\big) +
R\big(e_3R(2e_2)\big)
                                                     -R\big(e_2R(e_2R(2e_2))\big),\\
\beta_{16}&=& -R(e_6) + R\big(e_2R(e_4)\big) +
R\big(e_3R(e_3)\big)
                                                   +R\big(e_4R(e_2)\big)\\
          & & -R\big(e_2R(e_2R(e_3))\big) -R\big(e_2R(e_3R(e_2))\big)
                                               -R\big(e_3R(e_2R(e_2))\big)
                                                  +R\big(e_2R(e_2R(e_2R(e_2)))\big),\\
\beta_{24}&=& -R(e_3) + R\big(e_2R(e_2)\big),\\
\beta_{25}&=& -R(e_2^2) + R\big(e_2R(2e_2)\big)=R(e_2)R(e_2),\\
\beta_{26}&=& -R(e_4) + R\big(e_2R(e_3)\big)+R\big(e_3R(e_2)\big)
                                                 -R\big(e_2 R(e_2 R(e_2))\big),\\
\beta_{35}&=& -2R(e_2),\\
\beta_{36}&=& -R(e_3)+R\big(e_2R(e_2)\big),\\
\beta_{46}&=& -R(e_2), \ {\rm{ and}}\ \beta_{56}= 0.
\end{eqnarray*}}
The counter term expressions for the graphs $e_i$, $i=2,\dots,6$
are given in the first row of $\beta$,
$$
  \left < \phi_-\right |= \left < 1_\mathcal{T}\right | \beta
                        = (1, -R(e_2), -R(e_3)+R(e_2R(e_2)),\beta_{14},\beta_{15},\beta_{16}).
$$
These calculations for the $\beta_{ij}$ match the results of
Bogoliubov's counter term recursion applied to the coproduct
matrix. For the renormalized matrix character
$\widehat{\phi}\widehat{\phi}_{-}= \widehat{\phi}_{+}$ we find
\allowdisplaybreaks{
\begin{eqnarray}
  \widehat{\phi}_{+} &=& \exp\left(\tilde{\mathcal{R}}(\bar{\chi}(\widehat{Z}_{\phi}))\right)=
        {\bf{1}} - \tilde{\mathcal{R}}\left(\exp^{*_\mathcal{R}}
                       \left(-\bar{\chi}(\widehat{Z}_{\phi})\right)-{\bf{1}}\right)\nonumber
  \label{bogoliubov2}
\end{eqnarray}}
Then the renormalized expressions $\phi_+(e_i)$ for $i=2,\dots,6$
are obtained as components of the vector
$$
  \left< \phi_+\right| = \left < 1_\mathcal{T}\right | \widehat{\phi}_+
    = \left <1_\mathcal{T}\right | \widehat{\phi}\, \widehat{\phi}_-
    = \left < \phi \right | \widehat{\phi}_-.
$$
As a remark for the practitioner we mentioned that from
Proposition~\ref{inverse} we derive the more familiar equation for
the renormalized character \allowdisplaybreaks{
\begin{eqnarray}
  \widehat{\phi}_{+} &=& {\bf{1}} +
                               \tilde{\mathcal{R}} \big((\widehat{\phi}-{\bf{1}})\:
                                                                      \widehat{\phi}_{-}\big),
\end{eqnarray}}
which is just Bogoliubov's classical $\bar{\mathrm{R}}$-operation
giving the renormalized Feynman rules (matrix).


\subsubsection{Exponential approach of matrix factorizations}
\label{normal}

In this subsection we leave the realm of combinatorial Hopf
algebras of renormalization and come back to the results of
Section \ref{ss:matrixRB}, where we discussed the decomposition of
upper triangular matrices with entries in a commutative
Rota--Baxter algebra.

The following dwells on the exponential approach to the
calculation of the factor matrices $\bar{\alpha}_{\pm}$ in the
factorization of $\alpha$ in Corollary~\ref{matdecomp}. Other than
its theoretical significance, it also relates to the exponential
approach of the Birkhoff decomposition of Connes and Kreimer.

The reader should remember Theorem \ref{thm:RBalgMatrices}
asserting that with $(A,R)$ being a commutative Rota--Baxter
algebra, the triple
$\left(\mathcal{M}^u_n(A),\mathcal{R},\{\mathcal{M}^u_n(A)_k\}_{k\geq
1}\right)$ forms a complete filtered Rota--Baxter algebra.

Let us start with some properties of the exponential and logarithm
functions for complete filtered algebras of upper triangular
matrices defined in (\ref{expDef}) respectively (\ref{logDef}).
For any $1\leq n\leq \infty$, a natural basis of
$\mathcal{M}^u_n(A)$ is given by the matrices $E_{ij} \in
\mathcal{M}^u_n(A),\; 0 < i \leq j \leq n$, where the entry at
position $(i,j)$ is $1$, the rest zero. These matrices multiply
according to $ E_{ij}E_{kl}=\delta_{jk}E_{il}$. We can express the
logarithm and exponential in terms of this basis $\{E_{ij}\}$. For
$\alpha\in \mathfrak{M}_n(A)$, we have $\log (\alpha)\in
\mathcal{M}^u_n(A)_1$. So we have
$$
  Z_\alpha:=\log (\alpha)=(\tilde{\alpha}_{ij})
    =\sum_{0<i<j\leq n}\tilde{\alpha}_{ij} E_{ij}.
$$
These $\tilde{\alpha}_{ij}\in A$ are called {\it{matrix normal
coordinates}} (of the second kind). The concept of normal
coordinates in the context of Connes--Kreimer renormalization
theory appeared in \cite{Mexicans}.

For example, let the $3\times 3$ matrix $\alpha$ in
$\mathfrak{M}_3(A)$ be
$$
  \alpha= \sum_{i=1}^{3} E_{ii} + aE_{12}+bE_{13} + cE_{23}
$$
with $a,b,c \in A$. Note that $\alpha-{\bf{1}}$ is strictly upper
triangular and so $(\alpha-{\bf{1}})^k=0$ for $k\geq 3$.
Therefore, we have
\begin{eqnarray}
   \label{LieAlg}
 Z_{\alpha}&:=&\log (\alpha)= \alpha-{\bf{1}}-\frac{1}{2} (\alpha-{\bf{1}})^2
             = aE_{12} +  \left(b- \frac{1}{2} ac\right)E_{13} + cE_{23}
             \in \mathcal{M}^u_3(A)_1,
\end{eqnarray}
giving normal coordinates
$$
  \tilde{\alpha}_{12}=a,\ \tilde{\alpha}_{13}=b-\frac{1}{2}ac,\
  \tilde{\alpha}_{23}=c.
$$
Thus
\begin{equation} \label{example1}
  \alpha =
  \exp\left( aE_{12} +  \left(b- \frac{1}{2} ac\right)E_{13} + cE_{23} \right).
\end{equation}

In general, for given $\alpha=(\alpha_{ij}) \in \mathfrak{M}_n(A)$
these matrix normal coordinates can be calculated by the formula
\begin{equation}
 \tilde{\alpha}_{ij}=\alpha_{ij} + \sum_{k=1}^{j-i} \sum_{i<l_1< l_2 < \dots < l_k<j}
                                   \frac{(-1)^k}{k+1} \alpha_{il_1}\alpha_{l_1l_2} \dots
                                   \alpha_{l_k j}, \;\; 0< i < j\leq n.
 \label{normalcoord}
\end{equation}
These new coordinates allow us to write any $n \times n$-matrix
$\alpha \in \mathfrak{M}_n(A)$ as
$$
 \alpha=\exp\big(Z_{\alpha}\big)=
 \exp\left( \sum_{0<i<j\leq n} \tilde{\alpha}_{ij}E_{ij}\right).
$$

In order to obtain the factorization
$\alpha=\bar{\alpha}_+\bar{\alpha}_-^{-1}$, we need the
BCH-recursion (\ref{BCHrecursiono})
$$
 \bar{\chi}(Z_\alpha):=Z_\alpha -
 BCH\left(\tilde{\mathcal{R}}(\bar{\chi}(Z_\alpha)),
                      \mathcal{R}(\bar{\chi}(Z_\alpha))\right),
$$
which allows us to calculate $\mathcal{R}(\bar{\chi}(Z_{\alpha}))$
and $\tilde{\mathcal{R}}(\bar{\chi}(Z_{\alpha}))$ in
$\mathcal{M}^u_n(A)_1$. This is valid for any $n \leq \infty$.

We continue with our example of $\mathcal{M}^u_3(A)$. Note that
$\mathcal{M}^u_3(A)_1^k=0$ for $k\geq 3$ and that commutators of
higher order in $\bar{\chi}$ are identically zero in
$\mathcal{M}^u_3(A)_1$, due to the decreasing filtration. Thus by
(\ref{expFormulas}), we find \allowdisplaybreaks{
\begin{eqnarray*}
 \bar{\alpha}_{-}&=&\exp\left(-\mathcal{R}(\bar{\chi}(Z_{\alpha})\right) \nonumber\\
           &=& \exp\left(-\mathcal{R}\Big(Z_{\alpha} - \frac{1}{2}
                              [\tilde{\mathcal{R}}(Z_{\alpha}),\mathcal{R}(Z_{\alpha})]\Big)\right)\\
           &=& {\bf{1}} -\mathcal{R}\Big(Z_{\alpha} -
    \frac{1}{2} [\tilde{\mathcal{R}}(Z_{\alpha}),\mathcal{R}(Z_{\alpha})]\Big) +
                               \frac{1}{2}\mathcal{R}(Z_{\alpha})\mathcal{R}(Z_{\alpha})
\end{eqnarray*}}
We would like to underline, that up to this point we have not used
the condition that $\mathcal{R}$ is a Rota--Baxter map. Actually,
for the factorization, we only needed that
$\mathcal{R}+\tilde{\mathcal{R}}=\id_{\mathcal{M}^u_n(A)}$, which
is true for any linear map, and the BCH-recursion $\bar{\chi}$ (or
$\chi$) in a suitable topology, such as a complete filtered
algebra. The Rota--Baxter structure only enters in the next step,
when we replace the last term
$\mathcal{R}(Z_{\alpha})\mathcal{R}(Z_{\alpha})$ via Rota--Baxter
relation. Keeping in mind that, for the Lie bracket $[-\ ,-]$, we
have
$$
  [\tilde{\mathcal{R}}(Z_\alpha),\mathcal{R}(Z_{\alpha})] =
                 [Z_\alpha-\mathcal{R}(Z_\alpha),\mathcal{R}(Z_{\alpha})] =
                                             [Z_\alpha,\mathcal{R}(Z_\alpha)],
$$
thus
 \allowdisplaybreaks{
\begin{eqnarray}
 \bar{\alpha}_{-}
           &=& {\bf{1}} - \mathcal{R}(Z_{\alpha}) +
           \frac{1}{2}\big(\mathcal{R}(Z_{\alpha}\mathcal{R}(Z_{\alpha}))-
                         \mathcal{R}(\mathcal{R}(Z_{\alpha})Z_{\alpha})\big) + \nonumber\\
           & & \qquad\qquad\quad+ \frac{1}{2}\big(\mathcal{R}(\mathcal{R}(Z_{\alpha})Z_{\alpha})
                            +\mathcal{R}(Z_{\alpha}\mathcal{R}(Z_{\alpha}))
                                                    - \mathcal{R}(Z_{\alpha}Z_{\alpha})\big)\\
           &=& {\bf{1}} - \mathcal{R}(Z_{\alpha}) + \mathcal{R}(Z_{\alpha}\mathcal{R}(Z_{\alpha}))
                    - \frac{1}{2}\mathcal{R} (Z_{\alpha}Z_{\alpha}).
\end{eqnarray}}
In matrix form we therefore get
\allowdisplaybreaks{
\begin{eqnarray}
 \bar{\alpha}_{-}\!\!\!\!&=&\!\! \!\!\!\!\left(
        \begin{array}{ccc}
         1 & -R(a) & \big(-R(b)+\frac{1}{2} R(ac)\big)\\
         0 & 1 & -R(c)\\
         0 & 0 & 1
        \end{array}
  \right)
  +
 \left(
        \begin{array}{ccc}
         0 & 0 & R(aR(c))\\
         0 & 0 & 0\\
         0 & 0 & 0
        \end{array}
  \right)
  -
\left(
        \begin{array}{ccc}
         0 & 0 & \frac{1}{2} R(ac)\\
         0 & 0 & 0\\
         0 & 0 & 0
        \end{array}
  \right)\nonumber\\
  &=&\!\!\!\!
  \left(
        \begin{array}{ccc}
         1 & -R(a) & \big(-R(b) +  R(aR(c))\big)\\
         0 & 1 & -R(c)\\
         0 & 0 & 1
        \end{array}
  \right)
\end{eqnarray}}
The inverse of $\bar{\alpha}_-$ can be calculated, using the
recursive formula (\ref{antipodeInverse}) or directly from
(\ref{expFormulas})
\begin{eqnarray*}
 \bar{\alpha}^{-1}_{-}&=&\exp\left(\mathcal{R}(\bar{\chi}(Z_{\alpha})\right) \\
           &=& {\bf{1}} +\mathcal{R}\Big(Z_{\alpha} -
    \frac{1}{2} [\tilde{\mathcal{R}}(Z_{\alpha}),\mathcal{R}(Z_{\alpha})]\Big) +
                               \frac{1}{2}\mathcal{R}(Z_{\alpha})\mathcal{R}(Z_{\alpha})\\
     &=& {\bf{1}} + \mathcal{R}(Z_{\alpha}) + \mathcal{R}(\mathcal{R}(Z_{\alpha})Z_{\alpha})
                    - \frac{1}{2}\mathcal{R} (Z_{\alpha}Z_{\alpha})\\
&=&\!\! \!\!\!\!\left(
        \begin{array}{ccc}
         1 & R(a) & R(b)-\frac{1}{2} R(ac)\\
         0 & 1 & R(c)\\
         0 & 0 & 1
        \end{array}
  \right)
  +
 \left(
        \begin{array}{ccc}
         0 & 0 & R(R(a)c)\\
         0 & 0 & 0\\
         0 & 0 & 0
        \end{array}
  \right)
  -
\left(
        \begin{array}{ccc}
         0 & 0 & \frac{1}{2} R(ac)\\
         0 & 0 & 0\\
         0 & 0 & 0
        \end{array}
  \right)\nonumber\\
  &=&\!\!\!\!
  \left(
        \begin{array}{ccc}
         1 & R(a) & R(b) +  R(R(a)c)-R(ac)\big)\\
         0 & 1 & R(c)\\
         0 & 0 & 1
        \end{array}
  \right)
\\
  &=&\!\!\!\!
  \left(
        \begin{array}{ccc}
         1 & R(a) & R(b) -  R(\tilde{R}(a)c)\big)\\
         0 & 1 & R(c)\\
         0 & 0 & 1
        \end{array}
  \right)
\end{eqnarray*}

We similarly calculate $\bar{\alpha}_+=\exp\Big(
\tilde{\mathcal{R}}\big(\bar{\chi}(Z_\alpha)\big)\Big)$ and reach
the following factorization for example of $3\times 3$ matrix
$\alpha\in \mathfrak{M}^u_3(A)$.
\begin{equation}
\alpha= \bar{\alpha}_+ \bar{\alpha}^{-1}_- =\left(
        \begin{array}{ccc}
         1 & \tilde{R}(a) & \tilde{R}(b) - \tilde{R}(aR(c))\\
         0 & 1 & \tilde{R}(c)\\
         0 & 0 & 1
        \end{array}
  \right)
 \left(
        \begin{array}{ccc}
         1 & R(a) & R(b) -R(\tilde{R}(a)c)\\
         0 & 1 & R(c)\\
         0 & 0 & 1
        \end{array}
  \right)
\end{equation}
This recovers the factorization in (\ref{eq:barminuse}).\\

We finally make a remark on the normal coordinates for the example
in (\ref{bigexample}). For the matrix representation of the
character $\phi$,
$\Psi_A[\phi]=\widehat{\phi}=\exp\left(\widehat{Z}_{\phi}\right)$,
with $\widehat{Z}_{\phi}=\log(\widehat{\phi})$, the strictly upper
triangular matrix $\widehat{Z}_{\phi} \in \mathcal{M}^u_6(A)_1$
follows by using the formula for the matrix normal coordinates
(\ref{normalcoord}) \allowdisplaybreaks{
\begin{eqnarray}
\widehat{Z}_{\phi} &:=& \sum_{0 <i<j \leq 6} \tilde{\phi}_{ij}E_{ij}\\
           &=& \tilde{\phi}_{12}E_{12} + \tilde{\phi}_{13}E_{13} +
               \tilde{\phi}_{14}E_{14} + \tilde{\phi}_{15}E_{15} +
               \tilde{\phi}_{16}E_{16} + \nonumber\\
           & & \tilde{\phi}_{23}E_{23} + \tilde{\phi}_{24}E_{24}
                + \tilde{\phi}_{25}E_{25} + \tilde{\phi}_{26}E_{26} + \nonumber\\
           & & \tilde{\phi}_{34}E_{34} + \tilde{\phi}_{35}E_{35} +
               \tilde{\phi}_{36}E_{36} + \tilde{\phi}_{45}E_{45} +
               \tilde{\phi}_{46}E_{46} + \tilde{\phi}_{56}E_{56}    \nonumber\\[0.2cm]
           &=& \label{matZphi}
 \left(
        \begin{array}{cccccc}
   0  & \phi(\tilde{e}_2) & \phi(\tilde{e}_3) & \phi(\tilde{e}_4) & \phi(\tilde{e}_5)  & \phi(\tilde{e}_6) \\
   0  & 0                 & \phi(\tilde{e}_2) & \phi(\tilde{e}_3) & 0                  & \phi(\tilde{e}_4) \\
   0  & 0                 & 0                 & \phi(\tilde{e}_2) & 2 \phi(\tilde{e}_2)& \phi(\tilde{e}_3) \\
   0  & 0                 & 0                 & 0                 & 0                  & \phi(\tilde{e}_2) \\
   0  & 0                 & 0                 & 0                 & 0                  & 0         \\
   0  & 0                 & 0                 & 0                 & 0                  & 0
  \end{array}
           \right)
 \; \in \mathcal{M}^u_6(A)_1.
\end{eqnarray}}
We used the fact that $\phi \in G_A$ is a character, i.e. an
algebra homomorphism. We have the following simple polynomial
expressions for $\tilde{e}_i$, $i=1,\dots,6$, following from
(\ref{normalcoord})
\allowdisplaybreaks{
\begin{eqnarray}
 \tilde{e}_2 &=& e_2, \quad \tilde{e}_3= e_3 - \frac{1}{2}e_2e_2 \\
 \tilde{e}_4 &=& e_4 - e_2e_3 +\frac{1}{3}e_2e_2e_2, \quad
 \tilde{e}_5 = e_5 - e_2e_3 + \frac{1}{6}e_2e_2e_2 \\
 \tilde{e}_6 &=& e_6 - e_2e_4 -\frac{1}{2} e_3e_3 + e_2e_2e_3 - \frac{1}{4}
 e_2e_2e_2e_2.
\end{eqnarray}}
These are exactly the rooted tree normal coordinates as they
appear in \cite{Mexicans}, and \cite{FG,FGV}. They can be
calculated as well using the convolution product and the
logarithmic map,
$\tilde{e}_i=\log^{\star}\big(\id_{\mathcal{H}_{\mathcal{T}}}\big)(e_i)$,
$i=1,\dots,6$.\\

We hope that these examples provided some insight into the
underlying structure and calculational simplicity of the matrix
factorization in the context of complete filtered Rota--Baxter
algebras.\smallskip

Let us briefly summarize what we have found in this section. Upper
triangular $n \times n$ matrices, for any $1\leq n\leq \infty$,
with entries in a commutative Rota--Baxter algebra $(A,R)$,
$\mathcal{M}^u_n(A)$ form a complete filtered Rota--Baxter algebra
$\left(\mathcal{M}^u_n(A),\mathcal{R},\{\mathcal{M}^u_n(A)\}_{k
\geq 1}\right)$. The complete filtration allows us to define a
Baker--Campbell--Hausdorff based recursion relation, denoted by
$\bar{\chi}: \mathcal{M}^u_n(A)_1 \to \mathcal{M}^u_n(A)_1$, which
in turn gives rise to a decomposition of the group
$\mathfrak{M}_n(A)$ of upper triangular matrices with unit
diagonal. The linear map $\mathcal{R}$ appearing in the definition
of the recursion for $\bar{\chi}$ can be any linear map in
$\End(\mathcal{M}^u_n(A))$. Choosing it to be a Rota--Baxter map
gives rise to solutions of the matrix group factorization in terms
of recursion equations, Theorem~\ref{thm:main}.


\subsection{Berg--Cartier's ansatz using the grafting operation
on rooted trees} \label{sect:Berg-Cartier}

Berg--Cartier \cite{BergCartier} used a different but related
approach to encode the derivations in
$\mathcal{H}^*_{\mathcal{T}}$ in terms of lower triangular
matrices. For this they made use of the pre-Lie insertion
operation on Feynman graphs.

For the sake of convenience, let us state briefly the definition
of pre-Lie algebra. Let $A$ be a not necessarily associative
$\mathbb{K}$-algebra. We denote the multiplication $m_A: A \otimes
A \to A$ in $A$ by concatenation, $m_A(a \otimes b)=a\ b$, $a,b
\in A$. The associator is defined as $(-,-,-)_A:A\times A\times A
\to A$,
\begin{equation}
 (a,b,c)_A := a\ (b\ c) - (a\ b)\ c
\end{equation}
for $a,b,c \in A$. For $A$ being an associative
$\mathbb{K}$-algebra we have $(a,b,c)_A=0$ for all $a,b,c \in A$.

A (left) pre-Lie $\mathbb{K}$-algebra $(P,\diamond)$ is a
$\mathbb{K}$-vector space $P$, together with a bilinear pre-Lie
product $\diamond: P \times P \to P$, fulfilling the (left)
pre-Lie relation
\begin{equation}
 \label{preLie}
     (a,b,c)_P=(b,a,c)_P,\;\;\; \forall a,b,c \in P.
\end{equation}
or explicitly
$$
 a \diamond (b \diamond c) - (a \diamond b) \diamond c
  =  b \diamond ( a \diamond c) - (b \diamond a) \diamond c  ,\;\;\;
                                                \forall a,b,c \in P.
$$
The pre-Lie property is weaker than associativity, i.e. every
associative $\mathbb{K}$-algebra is evidently pre-Lie.  The
commutator $[a,b]:=a \diamond b - b \diamond a$ for $a,b \in P$
fulfills the Jacobi identity, making the $\mathbb{K}$-vector space
underlying $P$ a Lie algebra.

In the rooted tree setting the process of insertion of Feynman
graphs into other graphs becomes a grafting operation. The
derivations (\ref{Z}) $Z_t$ form a pre-Lie algebra $Z_{t'}
\diamond Z_{t''}:=\sum_{t \in \mathcal{T}} n(t',t'';t)Z_t$
\cite{CKI,CKII,CK4}, which by anti-symmetrization defines the
commutator of the Lie algebra $g$ of derivations (\ref{Liebra}).
This pre-Lie composition is used to define an action of the $Z_t$
on the vector space $\mathcal{T}$ freely spanned by the rooted
trees (or Feynman graphs). The operator representing the action is
denoted by $s(t):\mathcal{T}\to \mathcal{T}$, for all $t\in
\mathcal{T}$ and defined as follows
$$
  s_{t'} \left|t''\right> := \sum_{t\in \mathcal{T}} n(t',t'';t) \left|t\right>.
$$
As an example we calculate
$$
  s({\ta1}) \left|\ta1\right> = \big|\!\!\!\begin{array}{c}
                                            \tb2 \\
                                         \end{array}\!\!\big>, \qquad
  s({\ta1}) \big|\!\!\!\begin{array}{c}
                         \tb2 \\
                     \end{array}\!\!\big> = \left|\!\!\begin{array}{c} \tc3 \\ \end{array}\!\!\right>
                            + 2 \big|\!\!\!\begin{array}{c}
                                            \td31 \\
                                           \end{array}\!\!\!\big>,\qquad
  s({\tb2}) \left|\ta1\right> =  \left|\!\!\begin{array}{c} \tc3 \\ \end{array}\!\!\right>,\\
  \;\makebox {and by definition}\;
  s({t}) \left|1_\mathcal{T}\right>:=\left|t\right>\; \forall t\in \mathcal{T}.
$$
We used the ket-notation  for the rooted tree vectors introduced
in an earlier section. The rule for calculating the vector
$s(t)\left|t'\right> \in \mathcal{T}$ is to graft the tree $t$ in
all possible ways to the tree $t'$, and to multiply each tree in
the resulting linear combination by its symmetry factor. This
action can be used to define a representation of the $Z_t$'s in
terms of lower triangular matrices, which are just the transposed
of our upper ones (\ref{a}). The difference between lower and
upper triangular matrix representation reflects the fact that the
former increases the degree by grafting trees, whereas the latter
reduces them by ''elimination" of subtrees. The matrix
representation approach using the pre-Lie structure on Feynman
graphs or rooted trees appears to be limited to representations of
infinitesimal characters respectively characters. Relation
(\ref{linRep}) instead works for arbitrary elements in
$\Hom(\mathcal{H}_{\mathcal{T}},A)$.
\smallskip

It also appears that the ansatz chosen in \cite{BergCartier} for
the counter term matrix, denoted by $C_{1/\epsilon}$ in
\cite{BergCartier},
\begin{equation}
 \label{bcERROR}
 (C_{1/\epsilon})^{-1}=\exp\bigg(-\sum_{t\in\mathcal{T}}C(\phi(t)) s(t)\bigg),
\end{equation}
using the operator $s(t)$ for $t\in \mathcal{T}$ is not sufficient
for several reasons. First, the representation of characters $\phi
\in \Hom(\mathcal{H}_\mathcal{T},A)$ via the exponential map
demands the use of normal coordinates. Second, the counter term
character $\phi_{-}$, or its matrix representation
$\widehat{\phi}_{-}$, follows from the factorization of characters
in the sense of Atkinson, see Theorem (\ref{thm:Atkinson})
respectively Spitzer's identity for non-commutative associative
Rota--Baxter algebras, Theorem (\ref{complRB}). Therefore one must
include the particular properties of the Rota--Baxter relation as
well as the BCH-recursion (\ref{BCHrecursion}). This shortcoming
in \cite{BergCartier} becomes particularly evident when comparing
Equations (18), (19) and the one following (19), therein (see
Equation~(\ref{BCexample}) above for the correct expression).
Equation (18) is plagued with unwanted coefficients. The
derivation of the expression after Equation (19) in
\cite{BergCartier} is problematic as it seems to assume the
subtraction scheme map, denoted by $C$ in \cite{BergCartier}, to
be an idempotent algebra homomorphism. This is not true in
general, e.g. in the MS scheme $C=R_{ms}$, which keeps only the
pole part of Laurent series, is a projector of Rota--Baxter type.
Berg and Cartier earlier in their paper (page 18
in\cite{BergCartier}) proposed a special rule for products in the
image of the map $C$, e.g. $C(a)C(a)$ is supposed to read as
$\frac{1}{2}C(C(a)a)$, which in general is not sufficient to
resolve the aforementioned inconsistency. Instead, in a
commutative Rota--Baxter algebra $(A,R)$ we find
$\frac{1}{2}R(a)^2=R(R(a)a) - \frac{1}{2}R(a^2)$, $a\in A$.

The simple calculation of the counter term matrix (\ref{bcERROR})
used in \cite{BergCartier} would only apply to the subset of
ladder trees (or Feynman graphs) denoted by ${\ell d} \subset
\mathcal{T}$, if normal coordinates were properly
used\footnote{Cartier in a recent talk~\cite{Luminy} indicated an
approach, somewhat closer in spirit to the one presented here.}.
This follows from the fact that linearly ordered ladder trees form
a cocommutative Hopf subalgebra $\mathcal{H}_{{\ell d}} \subset
\mathcal{H}_{\mathcal{T}}$, or equivalently, the dually defined
derivations $Z_{t_{\ell n}}=:Z_n$, indexed by the number of
vertices of a ladder graph $t_{\ell n}$ form a commutative Lie
subalgebra $g_{\ell d} \subset g$, on which the BCH-recursion
reduces to the identity map, $\chi|_{g_{\ell d}}=\id$. Also, a
consistent motivation for reducing the algebraic Birkhoff
decomposition of the character matrix
$\widehat{\phi}=\widehat{\phi}_{+}\:\widehat{\phi}^{-1}_{-}$ into
a linear matrix-vector equation of the form (\ref{MatVectEq}) was
not given.


\section{Conclusion}

We have shown that the Connes--Kreimer Hopf algebra approach to
renormalization in pQFT can be entirely represented as a simple
and efficient triangular matrix calculus. Decomposing an $n \times
n$ upper triangular unital matrix as described above, provides us
with the counter term matrix as well as the renormalized matrix.
The matrix calculus allows for an efficient calculation of counter
terms, and henceforth renormalized Feynman amplitudes.\\
\smallskip

{\it{Acknowledgements}} We would like to thank
J.~M.~Gracia-Bond\'ia for useful and inspiring discussions.
I.~Mencattini provided helpful comments which improved the paper.
The second author thanks NSF and Rutgers University Research
Council for support. Dirk Kreimer is thanked for his corrections,
comments and remarks, especially with respect to
\cite{BergCartier}, and his patience with the first author while preparing
his PhD thesis.\\

\end{document}